# Grid tariff designs coping with the challenges of electrification and their socio-economic impacts

Philipp Andreas Gunkel[1*], Claire-Marie Bergaentzlé[1,] Dogan Keles[1,] Fabian Scheller[2] and Henrik Klinge Jacobsen[1]

[1]   Energy Economics and System Analysis, DTU Management, Technical University of Denmark, 2800 Kongens Lyngby, Denmark

[2]   Institute Zero Carbon (IZEC), Technical University of Applied Sciences Würzburg-Schweinfurt (THWS), Ignaz-Schön-Straße 11, 97421 Schweinfurt, Germany

*   Correspondence: phgu@dtu.dk;

**Abstract.** This paper investigates volumetric grid tariff designs under consideration of different pricing mechanisms and resulting cost allocation across socio-techno-economic consumer categories. In a case study of 1.56 million Danish households divided into 90 socio-techno-economic categories, we compare three alternative grid tariffs and investigate their impact on annual electricity bills. The results of our design consisting of a time-dependent threshold penalizing individual peak consumption and a system peak tariff show (a) a range of different allocations that distribute the burden of additional grid costs across both technologies and (b) strong positive outcomes, including reduced expenses for lower-income groups and smaller households.

**Keywords:** electricity grid tariffs; electrification; network cost distribution

## 1. Introduction

Policy initiatives such as the "Inflation Reduction Act" and the European "Green Deal" aim to reduce Greenhouse gas emissions with a particular focus on end-consumers (European Commission, 2019; U.S. Government Publishing Office, 2022). The part of the 2021 European Green Deal directed towards the electrification of heating systems and individual transport poses unprecedented challenges for European electricity systems. The European Commission targets the installation of 10 million heat pumps within five years until 2026 (European Commission, 2022), representing an additional 150 TWh of electrified heat demand. The European commission's "Fit for 55" proposed in 2021 effectively ends the sale of $CO_2$-emitting cars by 2035 (McPhie and Parrondo Crespo, 2022). Around 30 million electric vehicles are expected on European roads by 2030, increasing electricity demand by 84 TWh per year (European Commission, 2021). Both electric vehicles and heat pumps will increase demand by 100% or even 200% residential electricity demand in the coming decade leading to significant challenges to the existing electricity grid infrastructure (Andersen et al., 2021; Bollerslev et al., 2021; Systems and Group, 2021; Wangsness et al., 2021). European and US governments are well-engaged in creating incentive instruments for electrified end-use technologies. Consequently, investments are unavoidable to reinforce existing grid infrastructure and to respond to the upcoming demand boom (Clastres et al., 2019; Gautier et al., 2021). In particular, distribution grids face unprecedented challenges from incoming volumes of additional energy and peak effects driven by electrification. Distribution System Operators (DSOs) have a portfolio of tools to help the advanced operation of their networks, among which time-based [1]tariffs offer adequate means to address such generalized congestion problems as a whole (Cappers and Todd-Blick, 2021; European Commission, 2020). The question is how to design tariffs appropriately.

---

[1] Time-based tariffs refer to different structures varying in the temporal dimension like Time-of-Use (TOU) with fixed schedules, coincidental peak pricing and dynamic rates tied to short-term network usage, and hybrid TOU with the flexibility to invoke critical peak prices during exceptional congestion situations.

Recent literature on demand response has shed light on multiple impacts of price signals on system use and consumer flexible electricity use (Avau et al., 2021; Bergaentzlé et al., 2019; Cambini and Soroush, 2019; Fausto et al., 2019). Past literature also informs us about the potential risks but also windfall effects that tariff designs have on different groups of consumers with distinct socio-economic and technical characteristics stressing the challenges for vulnerable consumers (Azarova et al., 2018)

The literature on price signals for peak shedding or shaving discusses two main types of grid tariff designs for households that are either based on unit price differentiation over time or on introducing a power-based signal (Bjarghov et al., 2022a; Heleno et al., 2020; Hogan and Pope, 2017). Depending on the chosen scheme, the tariff design will either predominantly target energy use and support the reduction of system peaks or target power use and support the reduction of individual peaks (Council of European Energy Regulators, 2020). However, the current growth in electrical uses suggests better scrutiny of how to synthesize the attributes of both rate-making types. Because tariff designs affect consumer groups in a non-uniform way, it is also essential to look at the winners and losers of such a tariff. Additionally, it has to be considered that network usage is non-excludable but rival good allowing free-riding behavior if not adequately addressed (Abbott, 2001; Rubino, 2017).

This study designs and tests a new tariff design that limits system-wide congestion effects and effectively apportions system peak costs to the consumption that drives them. The design is compared to two tariffs targeting peak system reduction or individual peak reduction. In doing so, this study deconstructs the underlying fundamentals that govern the formation of these two tariffs: system peak and individual peak. Here, the key fundamentals to consider are the time variable (i.e. when do we choose to consider a period peak), and the power variable (i.e., above which grid capacity use do we consider the demand peak).

The first branch of the literature for electricity pricing in households shows that volumetric[2] grid tariffs with different time block rates are supportive of better grid usage (Bergaentzlé et al., 2019; Cambini and Soroush, 2019; Picciariello et al., 2015). The main goals have been to encourage demand response to reduce or shift peak consumption from the peak towards off-peak periods. In most cases in Europe, like in Denmark since 2019, or in the U.S., Time-of-Use tariffs setting pre-determined block rates were implemented (CEER, 2017; NordREG, 2015; TREFOR, 2022; Wangsness et al., 2021). While this type of signal allows for some degree of load smoothing, it is limited in sending flexibility signals in response to more critical events.

The shortcoming of fixed yearly schedules is tackled by introducing more dynamism in ToU setups. The literature mainly refers to critical peak pricing schemes (CPP) as a rate design enabling extraordinary peak signals to reflect system peak conditions (Faruqui and Sergici, 2013; Frontier Economics and Sustainability First, 2012). Empirically such tariffs trigger larger load shifts than simple ToUs due to the larger price spread between time blocks (CPS Energy, 2021a; Dütschke and Paetz, 2013; Faruqui et al., 2006). Furthermore, they are limited in the number of times they can be activated (10 to 15 days per year in the U.S. (CPS Energy, 2021b). Although the Critical Peak Pricing (CPP) tariff brings about more efficient utilization of the grid, increased flexibility, and reduced peak demand, it also presents a challenge in terms of allocating sunk costs without distinguishing between different grid users   (Council of European Energy Regulators, 2020).

Another branch of rate-making acknowledges the individual contribution to peaks, taking consumer load patterns and peak behaviors as a point of departure. Individual Peak Pricing (IPP) charges a higher tariff during peak consumption periods to penalize users with high peak load effects. Usually, this type of tariff applies a surcharge if the power demand is higher than a certain threshold. Such tariff has been tested or applied in various experiments and cases to constrain peak demand (Baldick, 2018; Zarnikau, 2014, 2013) but remains a poor indicator to coincide a customer's maximum demand with the system or local congestions (Borenstein, 2016; Hogan and Pope, 2017).

---

[2] Volumetric tariffs are pricing schemes that are based on volumes or quantities consumed over a determined period of time. In the context of many European system operators they are usually priced in e.g. €/kWh per hour.

Peak-coincident pricing is more suitable for solving congestion problems, associating the peak rate with consumptions occurring during system congestion hours (MIT Energy Initiative, 2016; Morell-Dameto et al., 2023). Abdelmotelleb et al. compare the response outcome of four different network charges, including the peak-coincident charge, showing that this design led to higher system economic efficiency (Abdelmotteleb et al., 2018). When applied to large industrial consumers with foreseeable peak load patterns, peak-coincident pricing drives is close to welfare optimizing behavior (Baldick, 2018). Azarova et al. test such tariff components on households showing that coincidental peak charges are the main factor for savings due to the random and short-term overlapping usage of several appliances (Azarova et al., 2018). However, the dataset (765 households) limits the scope of the analysis and only partially reveals how tariff designs affect categories of households. Peak-coincident pricing is efficient enough to apportion system costs across users in times of scarcity, but they do not distinguish between individual load contributions that effectively cause aggregated scarcity. Furthermore, while literature has covered optimal peak tariffs approximating them to forward looking long term marginal cost (Morell-Dameto et al., 2023), the cost distributional effects among consumers and the pathway towards them has not been in detail.

Time-based volumetric tariffs take their point of departure into system conditions reflecting system peaks, while individual pricing, whether coincidental or not, departs from individual peak loads. However, existing literature on grid tariffs neglects to investigate the notion of "peak" itself. To our knowledge, there is no clear definition of what a peak is, or rather when a system or household's demand is considered in a peak state. The closer a system operates on the technical boundaries, the likelier it is in a peak state. An aggregated system peak is the sum of all individual contributions, while some individuals use more capacity than others. Most time-based volumetric tariffs, however, treat each individual contribution the same. Consequently, the pricing mechanism in time-based volumetric tariffs treats the potential exclusion of certain grid users due to limited grid capacity in a uniform manner through marginal pricing. This approach doesn't differentiate adequately between individuals who contribute significantly to the scarcity and those whose contributions are comparatively lower.

The notion of using 5% to define system peaks in the U.S. and Europe has been well established in the policy literature, thanks to the work of Faruqui et al. (Faruqui et al., 2010, 2007). Koranyi justifies this threshold by explaining that the 5% corresponds to approximately 400 hours during which 90% of the total installed capacity in the U.S. is utilized (Koranyi, 2011). Past studies also use peaker capacities on the supply side as a reference to define the number of hours when peak rates should apply (Milligan et al., 2017). Many time-based volumetric tariffs build upon the load duration curve as their foundation. They achieve this by designating a subset of hours that surpass a specific aggregated installed capacity threshold. This subset of hours is used to symbolize the proportion of annual hours linked with peak periods.

The notion of threshold is also relevant at the individual level. In this case, the underlying question becomes how to define individual peak usage, which comes down to deciding what differentiates a "normal" consumption behavior from a peak behavior. Concretely, a consumer's maximum capacity threshold is limited by her physical capacity connection to the grid. Nevertheless, suppose each consumer connected to the same line can consume up to the limits of their individual physical capacity. In that case, it is not true that they can all do so simultaneously. Fausto et al. and Pérez-Arriaga et al. suggest symmetric pricing varying across consumers dependent on the state of the grid but also factoring in varying contributions to the aggregated peak (Fausto et al., 2019; Pérez-Arriaga et al., 2017). This may lead to a volumetric grid tariff level function that depends on individual consumption creating in practice hard-to-solve non-linearities[3].

We, therefore, suggest an approximation of this function in the form of a two-step dynamic approach that moves the pricing of aggregated scarcity and individual contributions to scarcity of capacity together while adding a dynamic temporal trigger. We introduce an hourly threshold consumption, which divides the base and peak tariff to reflect different grid capacity allocations per consumer while fragmenting and approximating the individual contributions to aggregated peaks in two levels. This study contributes to the state-of-art in questioning the main underlying assumption for variable ToU and individual peak pricing, which is the definition of peak period and peak level, respectively. We build on the two classic approaches of time-based volumetric tariffs and individual peak-coincidental tariffs and simulate different thresholds for aggregate peak periods duration in the former and individual peak thresholds in the latter. We finally develop a new tariff design at the crossroad between time-based volumetric and individual peak pricing to reduce both the system peak and individual peaks. The impact of these tariffs is comprehensively tested on a large sample of Danish households with various socio-economic characteristics and with or without an electric car or a heat pump. Our household dataset offers unprecedented detail by covering 1.56 million households divided into 90 different socio-techno-economic categories, including dwelling type and area, household income, occupancy, and electric vehicle or heat pump ownership.

Far from making a normative proposal on the qualification of the peak, this study explores in depth the redistributive effects related to the characterization of the individual peak and the characterization of the system peak. The results of this study provide a comprehensive overview of grid tariff designs and their impact on residential network bills, thereby offering system and network operators and regulators to understand cost-allocative effects. It does not aim to find one preferable solution but shows a menu of distributional effects of design approaches.

The key contributions of this study are:

- Developing time-varying volumetric grid tariff designs with a differentiated price mechanism that considers individual contributions to aggregated peaks differently.
- Investigating the impact of advanced electrification technologies, such as electric vehicles (EVs) and heat pumps, on the distribution of grid costs among households.
- Assessing the distribution of these time and quantity-based tariff designs among households belonging to different socio-economic groups.

The contributions of this study enable policymakers and network operators to gain a comprehensive understanding of how the transition from flat volumetric tariffs to time-based tariffs would impact the network bills of various consumer groups. This understanding empowers us to develop and propose novel grid tariff designs that address the complex relationship between individual and aggregated consumption. These designs aim to tackle the challenges associated with future constraints on network capacities while ensuring complete transparency regarding their

---

[3] To illustrate the non-linearity of grid tariffs from Fausto et al., consider the following dynamic grid tariff concept. Instead of relying solely on a fixed value tied to network conditions, e.g. in coincidence pricing, imagine a scenario where the tariff depends on both the network's state and how much each household consumes. This perspective leads to a cost formula for households like this:

$$Gt^{Peak}(Network_{status}, Consumption) * \ Consumption$$

In this equation, both the network's status and individual consumption play a role in the level of the individual grid tariff level. Due to consumption appearing in both parts of the formula that are multiplied together, the formula becomes nonlinear. However, it's important to recognize that while this idea is of theoretical nature, implementing it in the real world poses challenges and is thus not realistic.

effects on consumers. This study, only focuses on the part of the final retail electricity price of households that covers the volumetric distribution network tariffs.

This paper is organized as follows. Section 2 presents our methodology, encompassing the three grid tariff designs tested in this study and mathematical methodology. Section 3 presents the case-study context, the data and investigated scenarios. Section 4 presents the results which is followed by the discussion in section 5.. Section 6 concludes and offers policy recommendations.

## 2. Methodology

This section presents our approach to designing the three grid tariffs focusing on individual thresholds and system peak triggers and the combination of both peak definitions. Afterwards, we develop a pricing model for base and peak consumption by utilizing the definitions of these terms and applying the corresponding grid tariffs to calculate the annual grid bills by consumer groups and the total revenue of the system operator.

### 2.1. Grid tariff designs for individual and system peaks

The first design, Individual Peak Pricing (IPP), is presented in section 2.1.1 by defining a threshold that divides household consumption into base and peak consumption. In contrast, the second design presented in 2.1.2, Dynamic Critical Peak Pricing (DCPP), penalizes the critical timing of consumption. The third grid tariff design, Dynamic Critical Individual Peak Pricing (DCIPP) merges both approaches.

### 2.1.1. Defining individual peak power thresholds for Individual Peak Pricing (IPP)

The threshold divides individual consumption between peak and baseload. The lower the threshold, the more essential consumption, such as cooking or lighting, is defined as peak and subsequently subject to high rates. To address the ambiguity surrounding defining individual peaks and considering the variability in local technical characteristics, this study establishes four thresholds that penalize various types of consumption and, by extension, different user groups. We use sensitivity analysis to offer a menu of results to understand the dynamics of this definition on annual grid bills per consumer.

Figure 1 shows how much hourly household consumption varies between households with and without EV and heat pumps. While the lowest threshold of 1 *kWh/h* targets almost all the consumer's individual peaks, the range between 1 *kWh/h*, 1.5 *kWh/h* and 2 *kWh/h* targets particular heat pumps and electric vehicles. Advancing towards a 3 *kWh/h* threshold excludes most traditional consumer groups and heat pump users and defines electric vehicle charging as peak consumption in particular. We aim to observe the allocative effects of different levels by varying the threshold. Utilities can choose thresholds by allocating the available capacity among consumers.

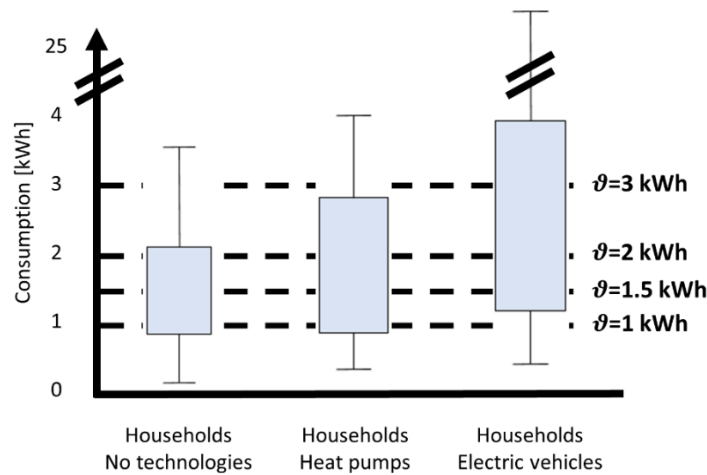

*Figure 1. Typical boxplot of hourly consumption by three different households with and without heat pumps and electric vehicles. The black dotted lines represent different threshold levels.*

The individual consumption of a group of consumers is divided by the following equation (1).

$$IPP: \quad q_g^{peak/base,year} = \begin{cases} q_g^{peak,year} = \sum_t^T q_{t,g} \quad if \quad q_{t,g} \geq \vartheta \\ q_g^{base,year} = \sum_t^T q_{t,g} \quad if \quad q_{t,g} < \vartheta \end{cases} \quad (1)$$

Where $q_g^{peak,year}$ is the consumption of the consumer group g classified as peak for every hour t if the consumption is above the threshold $\vartheta$. When the consumption is below the threshold $\vartheta$ we classify it as base consumption $q_g^{base,year}$. Figure A10 in the appendix A.1. gives a visual representation of when peak and base tariffs apply

IPP introduces a one-step differentiation of individual consumption to approximate different network usage levels and subsequently cost. As network usage is a rival good but non-excludable, the higher consumption of some individuals has potentially negative effects on other users (Rubino, 2017). IPP has therefore useful allocative characteristics. However, the design does not consider the temporal dimension of consumption. In many hours, higher consumption of some individuals has no adverse effect on others. This is the case for example in night hours which are often characterized by low residential consumption and thus offers households owning flexible technologies such as EVs and heat pumps the opportunity to shift their usage. Consequently, the following subsection introduces Dynamic Critical Peak Pricing (DCPP) that targets the temporal factor.

### 2.1.2. Defining system peak triggers for Dynamic Critical Peak Pricing (DCPP)

While the IPP design's challenge was defining a meaningful individual peak, the challenge of the DCPP is defining what is considered a meaningful peak at an aggregate level. In this study, the utility dynamically activates the DCPP periods when network capacities are scarce.

Defining the hours of critical network conditions is done via an approximation. As network conditions with high locational resolutions are unknown, especially at lower voltage levels, the Danish national load curve of 2017 is used as a proxy for the critical peak load situations. Figure A12 in Appendix A.1. visualizes the Danish load duration curve of 2017. A "**trigger percentage** $\theta$" divides the load duration curve into base and peak hours, as shown in Figure 2.

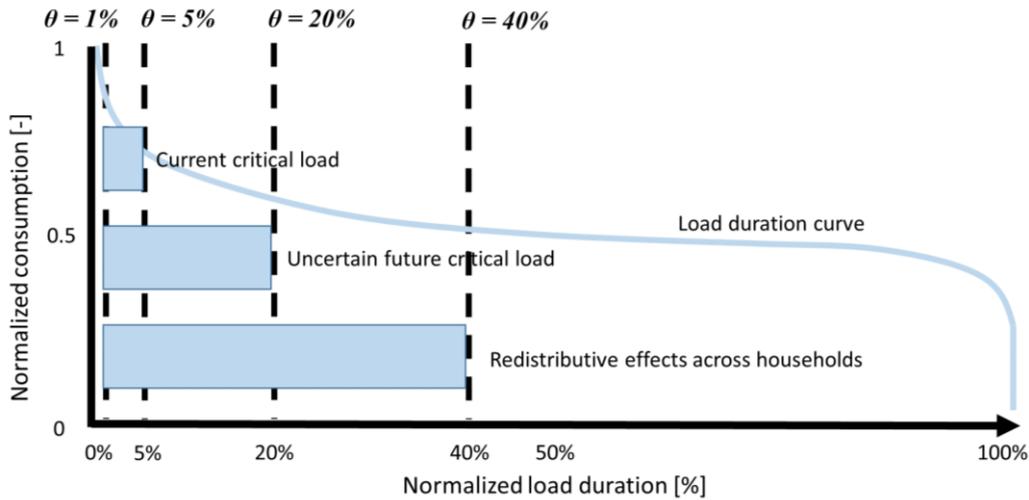

*Figure 2. Typical load duration curve and ranges of relevance regarding peak definitions and impact on the redistributive effects of different household categories. The black dotted lines represent different trigger percentage levels. Figure A12 in Appendix A.1. shows the shapes for 2017 in Denmark, which are common for the Nordic countries and are similar to other European industrialized countries.*

We acknowledge that the very definition of peak hours in the load duration curve will be subject to changes based on how fast green technologies penetrate the system. Our case study uses the top 1%, 5%, 20% and 40% of the Danish national load-duration curve of 2017 as the peak hours. Between 1-5% we assume the current critical load. With more adoption of flexible green technologies, it is not unlikely that the uncertain future critical load could go up to 20% (Gunkel et al., 2021a). Trigger percentages $\theta$ up to 40% serve as a sensitivity to understand redistributive effects better and provides a better overview of dynamics for policy following pricing rationale by the Danish Chamber of energy suppliers, Green Power Denmark (Dansk Energi, 2022). The more hours are defined as peak, the smoother and softer the redistribution across different consumer groups will be.

Equation 2 summarizes the division of peak and base hours.

$$DCPP: \quad q_g^{peak/base,year} = \begin{cases} q_g^{peak,year} = \sum_t^T q_{t,g} \;\; \forall \;\; t \;\; \epsilon \;\; H^{peak,\theta} \\ q_g^{base,year} = \sum_t^T q_{t,g} \;\; \forall \;\; t \;\; \epsilon \;\; H^{base,\theta} \end{cases} \qquad (2)$$

Hours that are within the top percentages defined above the peak threshold are defined in $H^{peak,\theta}$ and all consumption in those hours is defined as peak consumption. The residual hours that are in the lower range of the load duration curve within $H^{base,\theta}$ are defined as base consumption. Figure A11 in Appendix A.1. visualizes how peak and base hours are defined and grid tariffs apply.

DCPP introduces a temporal dimension and delivers a signal that triggers flexibility on every consumer when the system is congested. However, it misses allocative characteristics since all consumption levels by every consumer are treated equally. DCPP does not differentiate between peak driving users and consumers with minimal contribution, leading to a soft free rider effect coming from the previously named characteristics of networks being non-excludable but rival. The next subsection presents a design combining both tariff design characteristics to exploit both benefits.

### 2.1.3. Combining IPP with DCPP peak definitions: Dynamic Individual Critical Peak Pricing (DCIPP)

Dynamic Individual Peak Pricing (DCIPP) combines IPP and DCPP, and connects the allocative principles conveyed by the IPP with the flexibility of the DCPP. This scheme exclusively defines system peak hours when the hours of consumption are during system peak hours $H^{peak,\theta}$ and the individual consumption surpasses the threshold $\vartheta$ simultaneously. Equation 3 summarizes the categorization of peak and base consumption mathematically.

$$DCIPP: \quad q_g^{peak/base,year} = \begin{cases} q_g^{peak,year} = \sum_t^T q_{t,g} \quad \forall \quad t \; \epsilon \; H^{peak,\theta} \quad \wedge \; if \; q_{t,g} \; \geq \vartheta \\ q_g^{base,year} = \sum_t^T q_{t,g} \qquad else \end{cases} \quad (3)$$

As a result, DCIPP covers both design characteristics. At first, implementing the trigger percentage targets temporal flexibility and allows for high consumption when the network is not constrained. Second, higher grid tariffs are applied when the network is under stress to reduce congestion. DCIPP offers different characteristics by choosing thresholds $\vartheta$ and trigger percentages and thus offers to weigh both factors.

Figure 3 visualizes the application of the DCIPP tariff design on household consumption. Again, peak tariffs apply when consumption surpasses the threshold during hours of congestion. The peak tariffs are multiplied with the total hourly consumption, not just the excess. This is to cover the maximum distributional dynamics and, therefore, to answer the research question. The caption of Figure 3 contains an example of when and how the grid tariffs are calculated.

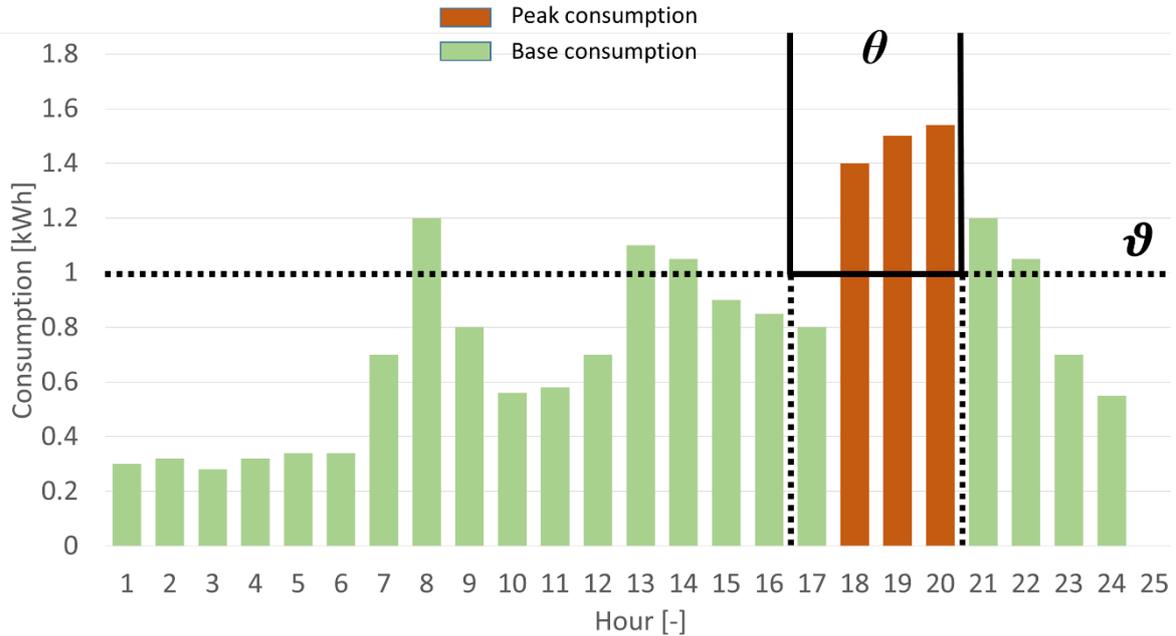

*Figure 3. Typical DCIPP grid tariff design applied to the individual household consumption pattern. Example in hour 18: The peak tariff is multiplied by 1.4kWh. Example in hour 17: The base tariff is multiplied by 0.8kWh.*

The investigations of DCIPP follow a sensitivity approach. All sixteen combinations of thresholds ($\vartheta$ = 1 $kWh/h$, 1.5 $kWh/h$, 2 $kWh/h$, 3 $kWh/h$) and trigger percentages ($\theta$ = 1%, 5%, 20%, 20%) yield different prominent effects on cost redistribution.

After categorizing the consumption of consumer groups into peak and base following the three design types, the mathematical formulation of the grid tariff model is presented.

### 2.2. Mathematical model to calculate grid tariff designs

The model for calculating the impacts of new grid tariff designs consists of a two-step formulation. Its central component is the constraint that the grid tariff retains the revenue-neutral total network income. The grid tariff with IPP, DCPP and DCIPP always consists of two parts: a base and a peak level. Before the optimization, the consumption of each consumer category is divided into the base consumption and peak consumption depending on the grid tariff design as an exogenous input. After that, the optimization determines the height of the IPP, DCPP and DCIPP peak tariffs. The following subsections present the constraints and initial values of the model.

The model calculates the total network income to be revenue-neutral to guarantee a clear view of the redistribution effects of the new grid tariff designs. Equations (4) and (5) summarize all the components of this constraint.

$$R^{SO} = \sum_g^G q_g^{peak,year} gt^{peak,var} + q_g^{base,year} gt^{base} \qquad (4)$$

$$gt^{base} = gt^{vol,2017} f^{recov} \qquad (5)$$

$R^{SO}$ represents the total network income earned by the system operator based on the original flat network tariff $gt^{vol,2017}$. This value is calculated once with a flat volumetric tariff without peak pricing to determine the sum and is kept constant for all other grid designs. The yearly consumption $q_g^{year}$ per consumer group G is divided into base consumption $q_g^{base,year}$ and peak consumption $q_g^{peak,year}$. The peak consumption is multiplied by the respective peak price $gt^{peak,var}$, the variable in this equation we are looking for. Similarly, the base consumption $q_g^{base,year}$ is multiplied by the base grid tariff, which is a constant input. The right-hand side of the equation, however, is also multiplied by a third factor $f^{recov}$. This represents the recovery factor for the base income. It is necessary to force the model to determine a difference between base and peak pricing and, therefore, to calculate a redistribution cost to maintain the total network income $R^{SO}$. With $f^{recov} = 1$ the variable $gt^{peak,var}$ has to equal $gt^{base}$ to maintain feasibility. This forces the model to find a $gt^{peak,var}$ that satisfies the revenue neutrality of the network's income and further results in the needed height of the peak tariff depending on the design. The level of $f^{recov}$ shall approach at least the long term marginal cost for investments into critical network elements (Dansk Energi, 2022; Morell-Dameto et al., 2023). At the same time, an immediate and significant change is not advised due to the potentially large cost redistribution for consumers and the subsequent public opposition. A step-wise approach for $f^{recov}$ to slowly approach the long term marginal cost is therefore implemented by DSO's in Denmark and this study therefore also includes the view on implementation pathways and their effect on consumers (Dansk Energi, 2022).

The redistribution factor changes the size of the redistribution but not the redistribution pattern across the different consumer groups. A sensitivity study on $f^{recov}$ therefore does not offer groundbreaking results and is consequently not shown in this study. The following results will only focus on the specific subpart of the retail prices relating to the volumetric grid tariffs. Relative changes for the entire retail bill would subsequently be smaller.

The next section presents the application of the model on a case study covering Denmark.

## 3. Case study: Grid tariff design in Denmark

System operators in Denmark have changed the grid tariff design after the smart meter rollout. This rollout further allowed for gathering vast amounts of data that can be used to perform analysis on a national scale. Thus, this section first introduces the dataset used in this study. After that, the values initializing the model are presented and lastly, the calculated scenarios are introduced. Appendix A.1. presents a small summary of the Danish context in grid tariff design.

### 3.1. Danish electricity consumption data and socio-techno-economic categories

This study uses the comprehensive electricity consumption dataset analyzed in (Andersen et al., 2021; Gunkel et al., 2021b) of 2017. Households are aggregated into different socio-techno-economic categories. The used categories are dwelling type and area, occupancy, income of the household, and the connection of green technologies such as EV and HP following the same approach of (Gunkel et al., 2021b). The categorization follows the recommendations of the European manual of statistics for energy consumption (European Union, 2013) to allow for socio-economic and policy investigations. Table 1 summarizes the division of the categories, using mainly median statistics. Theoretically, 288 potential categories could be present in this study, however, only 90 can be included due to GDPR requirements due to not enough data points within the categories. Ultimately, this study covers 1,565,856 Danish households of around 2,250,000 households.

*Table 1. Chosen socio-economic categories and their respective values.*

| Characteristic name | Characteristics | | | | |
|---|---|---|---|---|---|
| Dwelling type | | AP: Apartment | H: House | | |
| Occupancy | | P1: 1 occupant | P2: 2 occupants | P3: 3-4 occupants | P5+: 5 or more occupants |
| Dwelling area | AP: | A1<66$m^2$ | 66$m^2$<A2<85$m^2$ | 85$m^2$<A3 | |
| | H: | A1<110$m^2$ | 110$m^2$<A2<146$m^2$ | 146$m^2$<A3 | |
| Income level | | €1<240$kDKK$ | 240$kDKK$<€2<449$kDKK$ | 449$kDKK$<€3 | |
| Electric vehicle (EV) | | EV0: No | EV1: Yes | | |
| Heat pump | | HP0: No | HP1: Yes | | |

The house dwelling type (H) includes stand-alone single-family and terraced houses. The occupancy category contains single households (P1), two-person households (P2), 3-4 person households (P3) and households with five or more occupants (P5+). Due to their fundamental differences in size, the dwelling-area categories are separated for both houses and apartments. Since apartments are mostly smaller, the lower third (A1) goes up to 66 sqm, whereas the lower third of houses reaches 110 sqm (A1). (€1-€3) represent the income groups determined by median statistics. An EV in a household is indicated by (EV1), whereas an HP is represented by (HP1). For simplification, broader averages across several categories are presented in the results. We do not have information on storage systems or PV which had a negligible share in 2017 in the residential sector (Energistyrelsen, n.d.).

### 3.2. Initial values for the model and validation

Table 2 summarizes the initial values of the model. A representative grid area in Denmark was chosen, and its flat volumetric grid tariff for residential users (approx.. 0.025€) was applied for 2017 (Trefor, 2017). The calculated total network income results as $R^{SO}$ which is kept constant through all grid tariff designs and corresponds to approximately 102 m€. $f^{recov,base}$ is set to 1 as there is no redistribution between base peak consumption with flat volumetric tariffs.

*Table 2. Initial starting values of the model*

| Input | Value | Unit |
|---|---|---|
| $gt^{base}$ | 0.025 | €/kWh |
| $R^{SO}$ | 101,838,132 | € |
| $f^{recov,base}$ | 1 | - |
| $f^{recov}$ | 0.95 | - |

The resulting total network bills of all consumer groups with flat volumetric tariffs are used as a basis for all relative differences in the latter. In the main results, we then introduce new tariff designs with an adjusted recovery factor. The recovery factor of base consumption is set to 95% and therefore forcing the model to recover 5% of the total network income by the peak. This corresponds approximately to ToU tariff levels that we saw in 2019 in Denmark. $f^{recov}$ stays constant for the outcomes in the main result sections. The initial values have only a limited effect on the redistribution of cost. A higher or lower initial value of the grid tariff or a different recovery factor only changes the magnitude of the design impact and cost redistribution. In contrast, groups are positively or negatively affected by their consumption patterns and the respective grid tariff design. All relative results in the result section are calculated with the outcomes from the flat volumetric tariffs.

### 3.4. Scenarios and comparison of results

Overall, over 406 different scenarios were calculated for IPP, DCPP and DCIPP. This study presents twelve representative scenarios with four variations for each of the grid tariff designs for the sake of parsimony. Table 3 summarizes the scenarios shown in the results section.

*Table 3. Overview of presented scenarios*

| Scenario | Threshold | Trigger percentage |
|---|---|---|
| IPP; 1kWh | $1 kWh/h$ | - |
| IPP; 1.5kWh | $1.5\ kWh/h$ | - |
| IPP; 2kWh | $2\ kWh/h$ | - |
| IPP; 3kWh | $3\ kWh/h$ | - |
| DCPP; 1% | - | 1% |
| DCPP; 5% | - | 5% |
| DCPP; 20% | - | 20% |
| DCPP; 40% | - | 40% |
| DCIPP; (2kWh;1%) | $2\ kWh/h$ | 1% |
| DCIPP; (2kWh;5%) | $2\ kWh/h$ | 5% |
| DCIPP; (2kWh;20%) | $2\ kWh/h$ | 20% |
| DCIPP; (2kWh;40%) | $2\ kWh/h$ | 40% |

The IPP design threshold varies between 1-3 $kWh/h$ to show the sensitivity and allocative impact of different socio-economic households and technologies. DCPP scenarios vary across the trigger percentage between 1-40%, representing different pressure situations of the grid and also having allocative properties. DCIPP has both threshold and trigger percentages. The threshold is kept constant at 2 $kWh/h$ to simplify the result section, while the trigger percentage varies from 1-40%. The results section focuses on the relative difference in total volumetric grid tariff costs compared to the flat volumetric tariffs in 2017. The resulting prices of $gt^{peak,var}$ are not the main focus of this study because the level of the peak tariff is dependent on the local grid characteristics managed by the respective DSO. The peak tariff could vary across distribution systems in Denmark. However, the tendencies of the impacts on household categories and technologies stay the same.

In the next section, we compare the tariff designs presented in Table 3 to the residential grid tariff that was in effect during our consumption data collection. Danish households were charged a two parts tariffs based on a subscription fee and a fixed volumetric part applying a uniform fee to each consumed kWh (Dansk Energi, 2016).

## 4. Results

This section compares how the different tariff designs affect consumer groups' yearly electricity grid bills compared to uniform charges. The first subsection presents the effect of IPP and DCPP on households with and without HP and EV to illustrate basic dynamics. Thereafter, the result of DCIPP are visualized and described. The outcomes focus on the effects of DCIPP on consumer groups with different dwelling types, dwelling sizes and green technologies on yearly grid bills. Lastly, we highlight the redistribution of grid cost among different socio-techno-economic groups such as occupancy, income and technologies.

The resulting peak prices $gt^{peak,var}$ are summarized in Table A4 in Appendix A.2. and are occasionally referred to but are also presented in the respective subsections. The total redistribution of all ninety categories is summarized in Table A5 in Appendix A.2. Relative differences are calculated as the division of the total grid cost per category from the new design and the total grid cost from the flat volumetric tariffs as described in section 3.2..

*4.1. Comparing IPP and DCPP*

Figure 4 shows the relative difference in the yearly grid bill generated by the IPP to flat volumetric tariffs for households not owning an EV or HP (No-tech), with EV and with HP, respectively. The applied thresholds of IPP are between 1 *kWh/h* and 3 *kWh/h*.

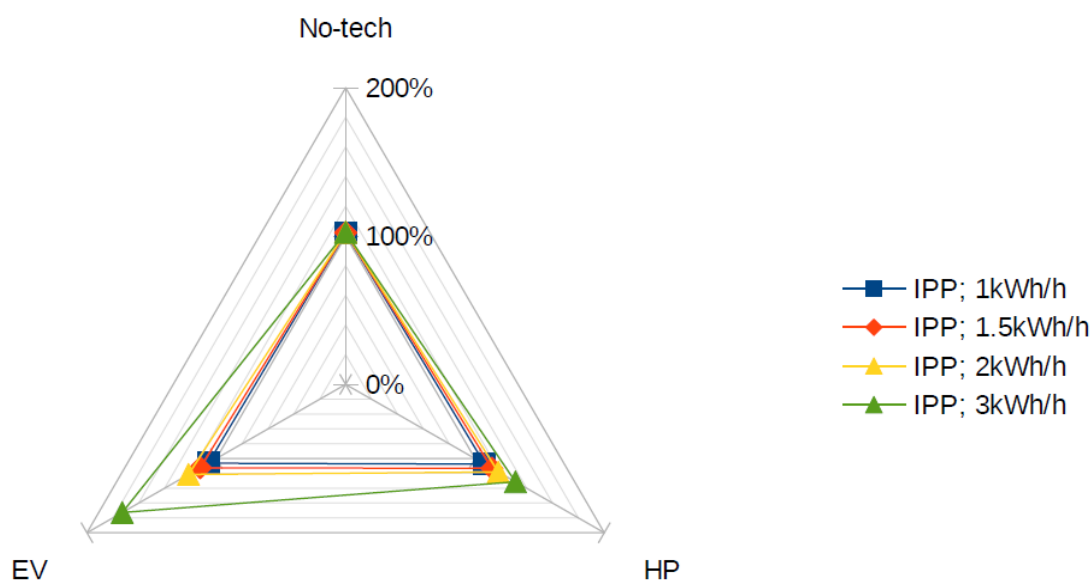

*Figure 4. Average relative redistribution of volumetric grid cost with IPP for a single-family house (H) with an occupancy of 3-4 (P3) in a high dwelling area (A3) and income group (€3) owning an HP, EV or not owning either.*

Generally, the higher the threshold, the higher the percentage change for HP and EV owners. Depending on the applied threshold, HP owners pay an additional 7.56% or up to 31.13%. EV users face an increase of up to 72.75% in volumetric grid costs, which can correspond to ~180 €/year more see Table A5 in Appendix A.2. Also noticeable is the effect of increasing the thresholds on technologies. 1 *kWh/h* and 1.5 *kWh/h* result in a 7.56% and 13.16% increase in yearly volumetric grid costs for HPs, whereas EVs yield only 6.26% and 12.49%. At the highest threshold, the picture shifts drastically toward the higher penalization of EVs. At the same time, the situation remains almost unchanged for households without technologies. Mostly EVs and a few large households (see Figure A13 in the appendix) or HP owners consume more than 3 *kWh/h* regularly, and therefore the main part of the DSO's peak income is recovered by them($q_g^{peak,year} gt^{peak,var}$). In contrast, with a 1 *kWh/h* threshold, the redistribution is relatively flat across user groups.

IPP reduces grid costs for smaller households with lower occupancy rates at the expense of larger households with higher occupancy (see Figure A13 in the appendix). Furthermore, the grid tariff design isolates the effect of high consumption, thereby increasing the grid costs of EVs in particular compared to HP. However, the design of IPP is inherently flawed due to its permanent application, which does not reflect the dynamics of congestion effects.

The variability of the dynamic congestion is tackled with DCPP. Figure 5Figure 4 shows that the triangle strongly shifts towards HP owners. A significant difference compared to the IPP design in Figure 4.

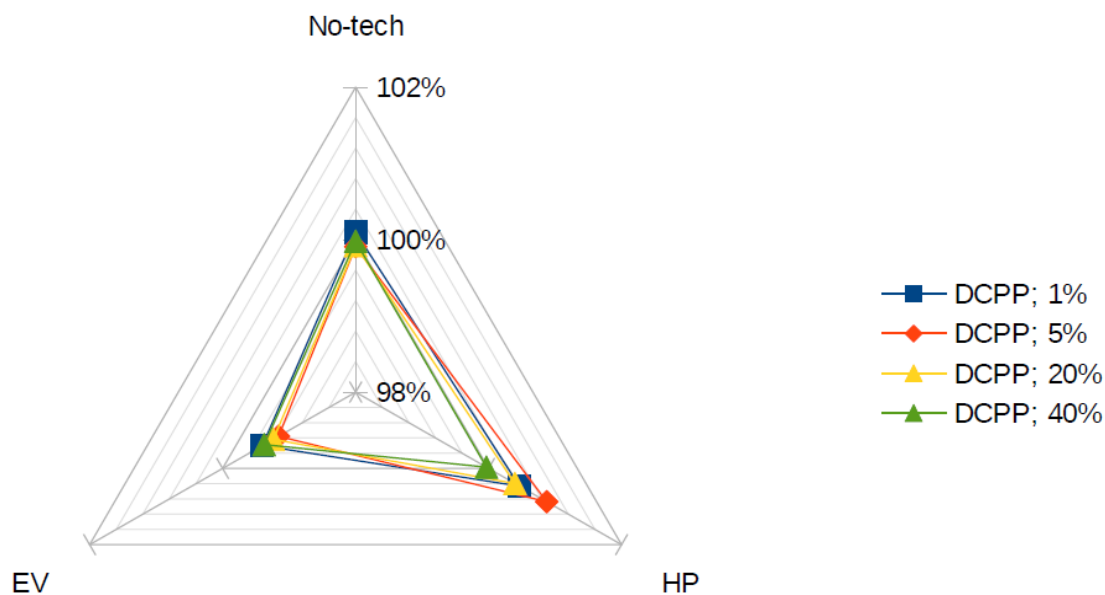

*Figure 5. Average relative redistribution volumetric grid cost with a single-family house (H) with an occupancy of 3-4 (P3) in the high dwelling area (A3) and income group (€3) owning an HP, EV or not owning either.*

The new design yields for the HP respectively 0.47%, 0.87%, 0.4% and 0.03% for the 1%, 5%, 20% and 40% DCPP scenarios. Generally, differences in yearly grid costs are almost negligible. The reason for that is the low recovery factor with $f^{recov}$=0.95 as introduced in section 2.2. to maintain comparability. However, differences become more prominent with smaller recovery factors (see also Table A7 and Table A6 in Appendix A.2.), and thus understanding the redistribution pattern is sufficient in this case. The reason for the smaller redistribution of grid cost lies in the consumption pattern of households. Only the share of consumption within the timing of the top percentages of national load determines the change in relative grid cost. At the same time, EV owners even save costs with DCPP. Their cost decreases by -0.6%, -0.84%, -0.76% and -0.62% in ascending order of all scenarios. The reduction of grid cost for EV owners is striking as they are considered one of the main drivers among heat pumps and PV for future grid expansion. Even though dynamic grid tariffs did not apply in Denmark in 2017, EV charging happened more during the night and outside of peak hours, showing flexibility to a certain degree (Gunkel et al., 2021b). The 40% DCPP scenario includes more hours of EV charging. Subsequently, the relative savings for EV owners decrease compared to the 20% DCPP scenario. Conversely, HP owners reach their highest relative cost increases in the 5% and 20% DCPP since HP consumption is exceptionally high during hours of high national load (Andersen et al., 2021; Gunkel et al., 2021b). As TOU is similar to DCPP, EV owners will likely save grid costs compared to simple flat volumetric tariffs. This may result in failing to apportion future grid reinforcement costs to the grid users with large peak-load patterns. At the same time, it is vital to acknowledge the already flexible behavior of EV owners.

Figure A13 and Figure A14 in appendix A.2 provide additional intuition for the effects on dwelling type and size. In general, both non-coincidental and coincidental peak tariff designs reduce yearly grid costs for smaller apartments at the expense of larger houses. At the same time, the magnitudes are different between both IPP and DCPP, dependent on $f^{recov}$. Further, the increase in cost is not necessarily happening smoothly, dependent on the dwelling type. For example, medium-sized houses are penalized higher than large houses (see Figure A14 in appendix A.2.). When we increase the recovery share from the peak by setting $f^{recov}$ to 0.9 and 0.8 we can further see that the relative difference are almost linear. Table A6 to Table A9 in the Appendix A.2. summarizes the sensitivity to the recovery factor. In the end, this means that the shape of the Figure 4 and Figure 5 maintain their shape even though we increase the recovery from the peak consumption.

DCPP offers efficient solutions to the problem of reflecting grid constraints dynamically without differentiating between consumption levels that contribute to congestion. DCPP reallocates grid costs that are less strong than IPP across consumers. In contrast, EVs pay significantly more with IPP. The DCPP design even reduces their grid expenses compared to flat volumetric tariffs at HP's expense, which pay more. The following section combines the advantages of the previously tested designs with the DCIPP.

## 4.2. DCIPP: reacting to network congestions and allocating network costs

DCIPP offers an opportunity to design the grid tariff freely. In particular, the triangle in Figure 4 and Figure 5 can be shaped based on preferences and previously named design objectives. Figure 6 and Figure 7 summarize the outcomes for apartments and houses of different dwelling areas and green technologies respectively.

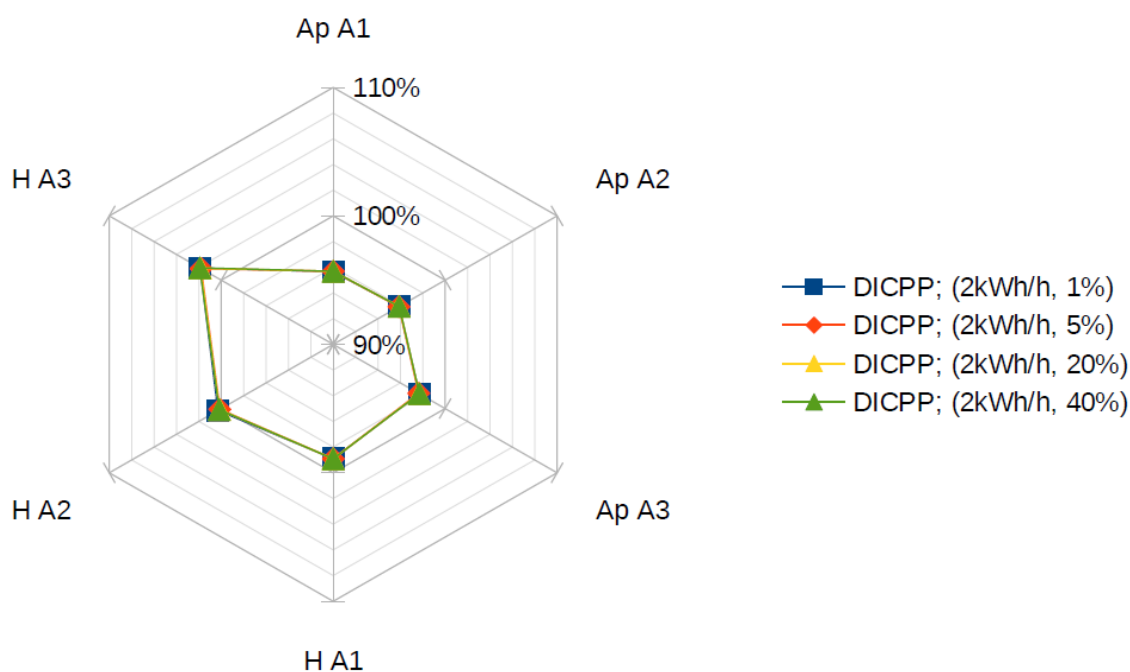

*Figure 6. Average relative redistribution volumetric grid cost with DCIPP for Apartments (Ap) and Houses (H) with different dwelling areas (A1-A3). No EVs or HPs are present in the households.*

Figure 6 shows that this arrangement of DCIPP has very stable impacts across the different scenarios. Differences between the scenarios with a fixed threshold (2 *kWh/h*) and varying trigger percentages (1-40%) range in second to third-digit percentages. Small, medium and large dwelling-size apartments save respectively 4.3%, 4.1% and 2.3% in yearly volumetric network costs. Small houses save about 1.1%, whereas medium houses pay up to 0.29% more. Large households pay, on average, a 1.85% higher grid bill. The minimal differences between the varied trigger percentages stem from the low consumption amount above 2 *kWh/h*. Consumers living in small apartments almost never consume more than the threshold. Consequently, increasing the number of peak hours (threshold) does not change their grid bill. Similarly, but to a lesser extent, the majority of consumption surpassing 2 *kWh/h* in large houses is already situated in the top 1%. Increasing the trigger percentage has therefore only a limited effect on them. Figure 7 visualizes the two groups with the largest grid cost redistribution.

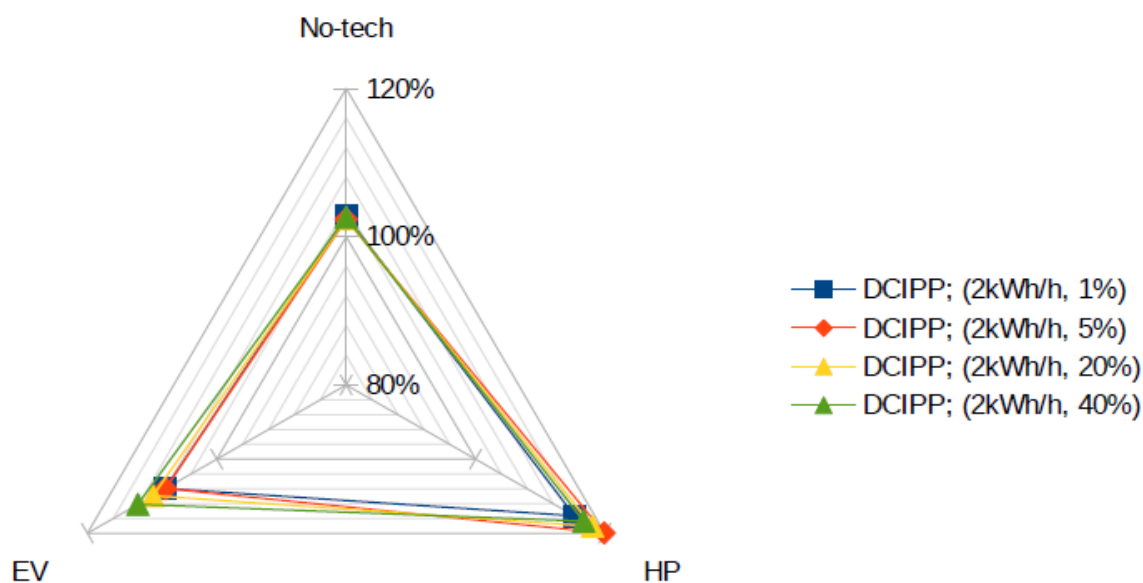

*Figure 7. Average relative redistribution volumetric grid cost with DCIPP for a single-family house (H) with an occupancy of 3-4 (P3) of the high dwelling area (A3) and income group (€3) owning an HP, EV or not owning either.*

Unlike Figure 4, where the tariff transfers more of the grid costs to the owners of electric cars, DCIPP has a more balanced impact on the owners of EV. At the same time, households without these devices are only marginally affected. Both technologies face higher costs. The triangle leans towards HP owners. With higher percentages included in DCPP part of DCIPP, the additional relative costs are 15.3%, 19.91%, 18.1% and 16.75% for each scenario. Similar to the outcomes of Figure 5, the 5% scenario yields the highest redistribution. Nevertheless, DCPP increases twofold the difference between the 1% and 5% scenarios, whereas the DCIPP results in less harsh developments. The highest cost increase is 57 €/year for a specific household category using a heat pump (see Table A5 in the appendix A.2.). Regarding EV owners, the effects of the scenarios are closer to the IPP, with steadily increasing costs. Ultimately, the four scenarios yield more minor additions of 7.91%, 7.94%, 9.86% and 12.21% compared to IPP. Saving potentials, or in other words, signals for shifting load from peak to base consumption, are also not equally high for different consumers. Households with EVs and heat pumps can save for every shifted % of peak consumption around 0.12-0.21% of their total cost, as can be seen in Table A10 in the appendix. Assuming a demand response of approximately 20% of their peak consumption, moving to base consumption leads to 48-51 €/year in their grid bill.

The DCIPP allows for a smooth redesign of the grid tariff. Small households benefit at the expense of larger ones. Depending on the chosen power threshold, the triangle of Figure 4 can be modeled relatively freely based on preferences. The higher the chosen threshold, the more the grid tariff penalizes the EV user. HP owners are stronger penalized between 5% and 20%. Triggering flexibility is only one of the possible preferences in designing grid tariffs. Another area of rising concern is redistributive effects on vulnerable households and in grid-cost allocation. The next section offers a detailed view of the tested tariffs' impact on the different income groups.

### 4.3. Limiting redistributive effects on poor households for an inclusive transition

This section zooms in on the effect of DCIPP on different income groups. Figure 8 shows the average relative cost change for households of all dwelling types, sizes and occupancies and for the three income groups €1 to €3, all without EVs or HPs. Figure 9 shows the households with technologies.

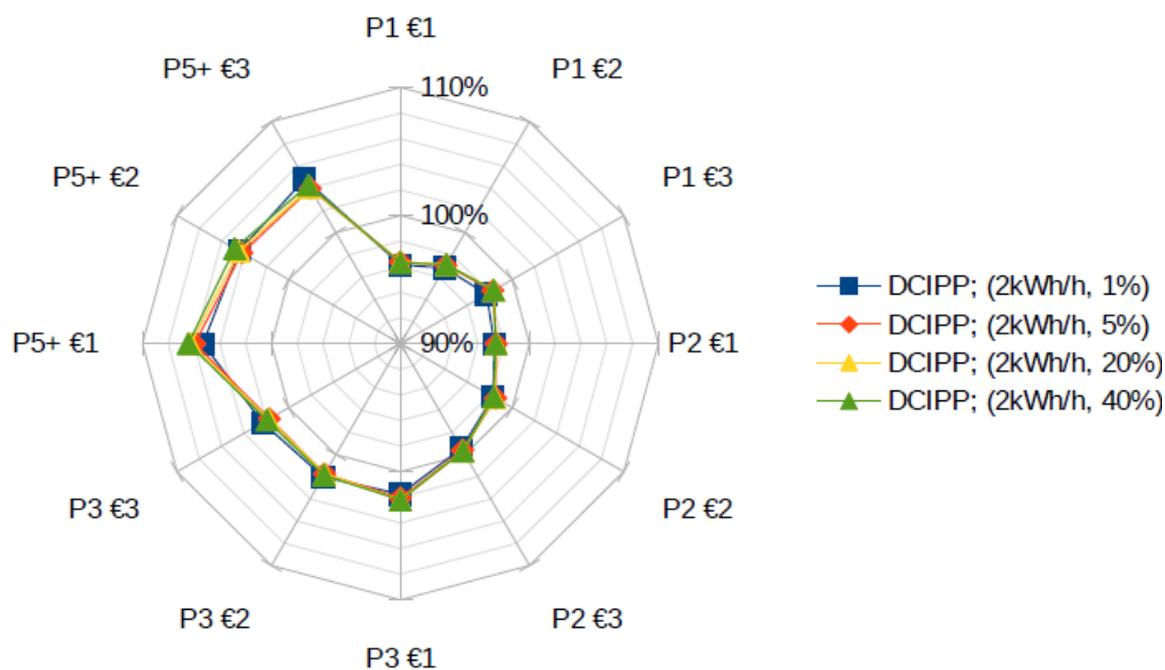

*Figure 8. Average relative change in cost with DCIPP for households without EVs or HPs across all dwelling sizes and types.*

The relative changes in grid-cost distribution are primarily linked to income and occupancy. Single households in the lowest income group benefit with around a 3.6% reduction in their average yearly grid bill. In the single household of the highest income group, the savings range between 2.28% and 1.66% depending on the scenario. The relative cost reduction shrinks with occupancy, turning into additional costs in the three to four-person households (P3) of all income groups. Large households with five or more occupants (P5+) show a small contrast, with the low-income group being more negatively affected than the high-income group. However, the poorer the household in the P5+ category, the higher the occupancy level (5.6 occupants in the low-income group against 5.1 occupants in the high-income group). Yet, this occupancy group faces 4% and 6.44% higher fees across all income groups.

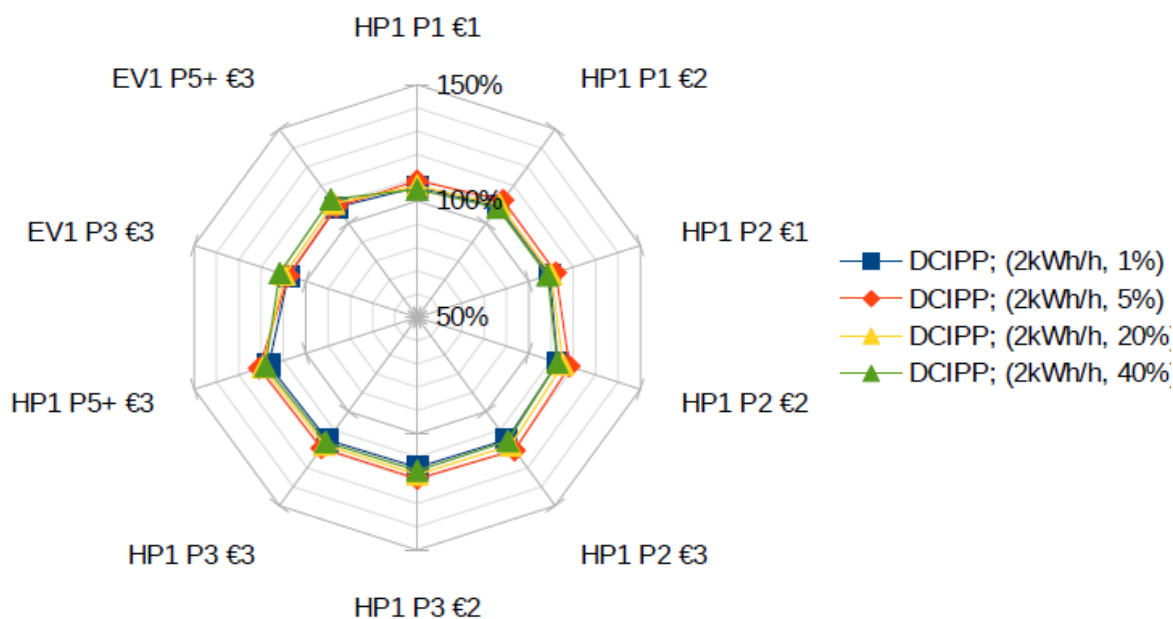

*Figure 9. Average relative change in cost with DCIPP for households with EVs and HPs across all dwelling sizes and types.*

Figure 9 adds to the income and occupancy the ownership of HPs and EVs. The more people living in a household, the more rooms are usually heated, and consequently, the larger the electricity consumption and grid tariff payments. A tendency to increase grid costs with higher incomes and occupancy is also visible in this figure. Only two-person households are present in all three income groups with an HP. Differences across income groups are present with a 12.1% over 18.25% to 20.61% rise in grid costs in the (2 *kWh/h*, 5%) scenario. The difference between the three to four occupancy compared to P5+ is marginal (up to 2% higher cost). EV households pay consistently less than HP households but still have around 10% higher grid costs than with flat volumetric tariffs. Figure A15-Figure A18 in Appendix A.3. summarize the impact on income groups of IPP and DCPP as shown in Figure 8 and Figure 9. The savings potential between different consumers has another dimension as it can also be seen as a price signal to shift load from peak to base as summarized in Table A10 in the appendix. Notably, singles living in apartments without any appliances see only marginal savings for reducing peak demand, which likely also requires more significant changes in behavior, such as shifted cooking times. Smaller price signals for those groups with lower or more pervasive flexibility potentials can be considered as an advantageous property. Contrastingly, large houses owning an EV or HP are facing higher signals, or in other words, savings potentials. Shifting only 20% of e.g. EV charging from peak to base hours can result in savings of around 48 € in the yearly grid bill under DCIPP. Savings from 20% EV charging under DCPP only yield a quarter of DCIPP savings at the expense of other household types.

In contrast to IPP and DCPP, the DCIPP design increases mostly volumetric grid costs the higher the income and occupancy of a household. This pattern is also maintained when adding technologies such as EV or HP to households. DCIPP offers to model a grid tariff design that distributes costs depending on preferences and maintains outcomes dependent on income level and occupancy.

## 5. Discussion

In the following, we first present a critical review of the assumptions and used method. Afterwards, we propose an in-depth discussion on grid tariff characteristics and how to design them.

*5.1. Critical review of the assumptions*

We rely on four key assumptions that affect our results. First, we assume that residential demand is inelastic to changes in network tariffs, which results in changes in yearly network bills. Consumers are generally cost-averse; therefore, increasing tariffs in specific periods triggers their demand response (Schofield, 2015), especially when also equipped with intelligent controllers (Avau et al., 2021; Nicolson et al., 2017). However, the flexibility of technologies and consumers cannot be directly assessed from the demand data. At the same time, EV consumption in this study already indicates a degree of flexibility, as is shown by an average reduction in grid costs when applying DCPP (see Figure 5). Consumers in Denmark could already decide on real-time spot-pricing in 2017, and shifted EV charging is thus present in the dataset (Frits Møller Andersen et al., 2021; Gunkel et al., 2021b). The direct reason for the shift in EV charging remains unknown. Thus, the interaction between the system operator and the consumer is dynamic. It requires yearly updates of grid tariff levels and potential designs, discussion of which would go beyond the scope of this study. Table A10 provides the only indication of potential savings for demand response.

There exists an extended literature intended at estimating the saving that such tariff types may generate that typically ranges between 3 to 30% peak load reduction, depending on multiple parameters such as the type of used electrical appliances, the height and duration of the peak period, the presence of communication devices, automation, and general information and education material, as summarized in (Bergaentzlé et al., 2014). Generally speaking, peak-specific signals trigger a larger peak reduction response compared to ToU pricing, but ToU tends to trigger lower withdrawn volumes over longer periods. How the pricing schemes translate into energy bill savings is highly contexts-specific and will mainly depend on whether the tariff is bundled or not and on the electricity and grid cost in time.

Further, an interaction, overlay, and correlation exist between electricity prices from the market and the grid tariffs that are difficult to disentangle. (Ansarin et al., 2020) Investigate the cross-subsidies of prosumers with different electricity pricing designs, including price elasticity of consumers excluding grid tariffs. Both signals are supposed to nudge the consumer to react to certain generation or network conditions. The overlay of electricity prices passing from the wholesale market and local grid tariffs signaling congestion gets more relevant in the future. While the adoption of real-time prices in Denmark gets more wide spread due to the energy crises (Ritzau, 2022), the grid component gets updated over time, driven by Danish DSO's (Dansk Energi, 2022). With the high penetration of renewable capacities in Denmark amounting to approximately 80% of zero-marginal price production of electricity, real-time prices might not be appropriate to signal scarcity for both generation and grid together which represents a methodological difference to (Ansarin et al., 2022) while the effective modeling of the consumers is arguably done in an innovative fashion and a shortcoming of this study with a focus on ex-post analysis (Norlys.dk, 2023).

Second, we use a proxy for grid congestion in peak hours given by the load duration curve. The definition of congestion in distribution systems is a local challenge. It depends on the local grid infrastructure, consumer behavior and connected technologies. Consequently, the investigated grid tariffs require as detailed a geographical and temporal resolution as possible (Fausto et al., 2019). Moreover, the level of the peak tariffs follow a pathway approach rather than using the long term marginal cost for network critical network components as they are unknown for the specific case study (Morell-Dameto et al., 2023). The cascading effect on different voltage levels from flexible demand and consequently the impact on network usage and expansion requirements are not covered in this study but implicitly and potentially covered in the sensitivities of $f^{recov}$. At the same time, certain scarcity signals and their timing, e.g. from medium voltage levels, are not included in this study.

Third, this study only investigates a snapshot based on 2017 data and does not assume any changes brought about by adopting EV and/or HP. Adopting these technologies changes the number of households present in each category and the grid-cost distribution between them. Updating grid tariffs requires anticipating future developments across the parameters and categorizations. This analysis delivers a menu of different cost allocation mechanisms while still being efficient in handling efficient grid utilization. The role of distributed generation investigated in (Ahmad, 2021; Gautier et al., 2021) and (Cambini and Soroush, 2019) is further not touched upon in this study.

Finally, for reasons of simplicity, this study only focuses on a subset of twelve chosen scenarios among the 406 generated scenarios (showing different thresholds, trigger percentages, peak hours, one threshold and also two thresholds). The chosen scenarios are mostly edge cases to maximize the characteristics of the designs. Our findings indicate that the DCPP demonstrates superior efficiency in terms of grid operation as it targets all consumption with stronger signals leading to more demand response. However, transitioning to the DCIPP design primarily enhances cost allocative characteristics and the perception of better fairness dependent on the preferences of the regulators, potentially compromising overall efficiency.

### 5.2. Comparing grid tariff design characteristics

Although the main objective of time-varying tariffs is to take advantage of demand-side flexibility, tariffs should not overlook relevant socio-economic issues. Grid tariffs are regulated prices that underlay several objectives. System operators and regulators have to deal with the complexity of multiple objectives when choosing a design. With several overlaying objectives, it becomes implicitly challenging to balance the technical, economic and socio-economic factors, as they might cause opposing effects in the context of a rapidly changing environment. The new loads resulting from household electrification add and will continue adding to already existing electricity demand, potentially creating a competition effect across energy uses when the grid is congested. The higher grid charges triggered by, e.g., large simultaneous EV charging will also affect households not concerned by this charging, including the most vulnerable ones, those that only cover their basic needs. This local competition effect increases the relevance of including socio-economic considerations when designing grid tariffs.

DCIPP combines both IPP and DCPP to balance their characteristics in terms of timing and distinguishing individual contributions (discussed in depth in Appendix A.3.). DCIPP offers the possibility to support more equitable solutions for vulnerable households that use electricity for their basic needs while transmitting to a chosen degree efficient signals to the households causing congestions and peak system load, and associating grid-cost recovery with wealthier and larger households. The analysis shows that the higher the income in the same occupancy group, the more cost-sharing could be determined (see Figure 6).[4] Also, smaller flats and houses pay less than larger ones (regardless of occupancy), which mostly correlates with income. With DCIPP, households with higher incomes must pay slightly higher tariffs. In contrast, lower-income households benefit from this tariff design compared to the flat-tariff system that currently applies to most households. This characteristic is useful for regulators wishing to increase the acceptability of changes to grid tariffs with a consumer-centric approach, as highlighted by (Neuteleers et al., 2017). The differentiation of signals and savings summarized in Table A10 further indicates that consumers with higher flexibility potential and effect on the grid have the most pronounced targeting.

A similar tendency applies to households with heat pumps: households with higher incomes pay slightly more than those with a flat rate. However, we did not find a significant distributional effect among income groups in the category with households owning EVs. At the same time, based on our design, EV owners are less disadvantaged, even though they have high consumption. The reason for this could be that they mainly charge outside the times when the high tariff of the DCIPP applies.

---

[4] Only for large households (category P5+) does this seem to be different. However, this must be handled cautiously, as the different income groups within P5+ are not comparable. The average occupancy in the low-income group is 5.6, and in the high-income group 5.1. See Section 4.3.

The extent to which this design effectively shields vulnerable users against high grid costs depends on how the threshold for defining peak load is set and therefore refers to policy arbitrage, preferences and decisions and a definition of basic electricity needs of households (e.g. cooking and heating). The threshold at which excess consumption is classified as peak consumption and charged at a higher tariff must be set carefully. A potential challenge is that large households are likely to surpass a threshold that is dimensioned to target mainly households with EVs. A balanced threshold needs to be found so that the DCIPP is still effective in shifting demand to off-peak hours and, on the other hand, does not penalize the most vulnerable groups for allowing an inclusive and equitable transition. The threshold can potentially be set to available local capacities or local consumer characteristics and technologies in order to trigger grid-friendly behavior.

Finally, it can be summarized that DCIPP allows grid costs to be allocated across different consumers and technologies. In contrast to fixed charges based on e.g. technologies, this tariff design only targets the impact of residential consumption on the grid at moments of congestion rather than penalizing household connection to the grid in general. DCIPP also targets the challenge of cost allocation during congestion times, different from dynamic capacity tariffs (Bjarghov et al., 2022b). Dynamic capacity tariffs must be bought or auctioned for a chosen period. Choosing a good level is a complex risk-hedging task to solve for consumers. Dynamic volumetric tariffs can be seen as a weighted discretized derivation of capacity tariffs and, thereby, dependent on the chosen temporal fragmentation, reflecting the cost of consumption more precisely during the hour of congestion. Consumers pay more with DCIPP when the network is congested compared to capacity tariffs, where they pay when they might or might not use the grid during hours of congestion. The allocation of scarce capacities can further lead to competition effects between network users as wealthy households can overbid vulnerable consumers or policy instruments have to handle the equitable allocation of scarce network capacities. At the same time, adding a dynamic capacity tariff solves the challenge of limiting the capacity of consumers in moments of immediate threats of local blackouts by creating a pre-determined merit order of consumers. Ultimately, the DCIPP design is a balance that adds allocative features at the expense of fully penalizing consumption during peak periods.

Our proposed grid tariff designs have generalizability and relevance beyond the specific case study we applied them to. The EU has identified the need to update grid tariff designs to accommodate electrification as a key objective, and the cost effective implementation of these designs requires the use of smart meters, which is a target set by the EU Commission (The European Commission, 2014). While the rollout of smart meters has been slower than initially planned, the proposed grid tariffs are applicable across the EU as soon as the infrastructure is enrolled. Given the electrification of heat and transport in many EU countries, along with varying peak demand patterns and grid capacities, the challenges addressed by our proposed designs are applicable across the EU, albeit with different magnitudes and specific contextual factors to consider.

## 6. Conclusion

This study aims to provide a detailed view of grid tariffs for residential electricity consumption and their distributional properties disaggregated into the socio-techno-economic categories of residential households. In order to address several tariff design objectives, such as congestion management, allocative characteristics and potentially perceived fairness, we have developed and tested three definitions of individual and aggregated peaks and, subsequently, grid tariff designs from a consumer-centric view. First, Individual Peak Pricing (IPP) penalizes consumption above a certain threshold. Second, Dynamic Critical Peak Pricing (DCPP) penalizes consumption during hours of peak loads. Third, Dynamic Individual Critical Peak Pricing (DCIPP) combines the two. We calculate the redistribution of yearly volumetric grid costs under the assumption of the income neutrality of a distribution system operator that applies flat tariffs to around 1.56 million Danish households following the approach of (Azarova et al., 2018).

IPP increases the volumetric grid costs, mainly for large households and households owning electric vehicles and heat pumps. In contrast, small households benefit from this tariff by seeing a reduction of their grid bills. The higher the threshold of the consumption volume to trigger the peak tariff in the IPP design, the more EV owners face higher volumetric costs. IPP is poorly suited to reflecting grid costs in the temporal dimension, thus alleviating peak loads. DCPP, on the other hand, corrects this shortcoming, as it applies higher rates when the network is congested. The design reduces the cost for apartments at the expense of medium-sized houses. Heat pump owners face higher network bills due to their consumption patterns. Conversely, electric vehicle owners pay even less than with flat volumetric tariffs, suggesting a certain flexibility but a lower contribution to total grid-cost recovery in Denmark. It is therefore important to review the volumetric grid-cost allocation on households from current TOU designs as well, as they might not allocate cost as regulators potentially desire.

DCIPP combines both allocative and temporal characteristics, offering a balanced consumer-centric design. Moreover, system operators and regulators have the opportunity to balance the grid-cost allocation between electric vehicles and heat pump owners and incentivize either temporal or peak load shaving. The grid tariff design can also be adapted to local network capacities and consumer characteristics. DCIPP can also be adjusted to both strong and weak grids, district heating areas or areas with dominant electric vehicle charging patterns. Our study further shows that DCIPP has characteristics that mostly reduce network bills in the smaller and the poorer households. Although not the primary objective, the results of DCIPP regarding its impacts on household income groups offer regulators and system operators a tool for increasing the acceptability of new grid tariff designs to tackle future challenges.

The DCIPP therefore offers a useful tariff design for regulators and system operators to price network usage based on different preferences and local network characteristics. This is done by weighing the importance of allocative and fairness preferences against the characteristics affecting congestion management. While this study does not directly model demand response and cascading effects, it contributes to the existing literature by comprehensively analyzing various grid tariff designs. It examines their progression towards aligning with long-term marginal costs and their implications for grid bills from a holistic perspective of residential consumer groups. Furthermore, it introduces a cost-allocative element to the framework, enhancing its depth and scope. Future research endeavors should address the aforementioned limitations. Additionally, incorporating adjustments to peak tariffs based on forecasts of green technology adoption will be essential to establishing optimal congestion management."

## 7. Patents

**Author contributions.**

Philipp Andreas Gunkel: conceptualization, methodology, software, analysis, writing

Claire-Marie Bergaentzlé: conceptualization, writing, supervision

Fabian Scheller: conceptualization, methodology, supervision

Dogan Keles: methodology, writing, supervision

Henrik Klinge Jacobsen: methodology, writing, supervision

All authors have read and agreed to the published version of the manuscript.

**Funding.** This article was partly funded by the FlexSUS project (nbr. 91352), which received funding in the framework of the joint programming initiative ERA-Net RegSus, with support from the European Union's Horizon 2020 research and innovation programme under grant agreement No 775970. Fabian Scheller acknowledges the financial support of the European Union's Horizon 2020 research and innovation programme under Marie Sklodowska-Curie grant agreement no. 713683 (COFUNDfellowsDTU)

**Conflicts of interest.** The authors declare no conflict of interest. The funders had no role in the design of the study, in the collection, analysis or interpretation of data, in the writing of the manuscript, or in the decision to publish the results.

## Appendix A

We acknowledge that recent works inform on more advanced tariff designs for solving local grids based on dynamic capacity pricing (Bjarghov et al., 2022b; Neuteleers et al., 2017). Both show theoretically promising outcomes. However, despite receiving support in Europe from the European Agency for Energy Regulators (European Union Agency for the Cooperation of Energy Regulators, 2021), such tariffs remain seldom offered to households due to their complexity and the necessity to deploy high-level automation equipment (Council of European Energy Regulators, 2020; Fell et al., 2015; Numminen et al., 2022). Finally, experts anticipate that simpler tariffs such as ToU and CPPs will progress the most in the future (Faruqui et al., 2020).

### A.1. Method and context

Figure A10 visualizes an illustrative example for setting the threshold $\vartheta$ to 1 kWh per hour. When total consumption is below the threshold line, it classifies as base consumption $q_g^{base,year}$ (green bar), when it surpasses the line it classifies as peak consumption $q_g^{peak,year}$ (red bar). Classifying all consumption as $q_g^{peak,year}$ when it surpasses the threshold is a design choice to maximize the difference in order to explore the extreme implementation boundaries of the design. Grid tariffs are paid as follows for hour 18: 1.4kWh x peak grid tariff. The peak IPP tariff is represented by the red consumption bars, and the base grid tariff by the green consumption bars.

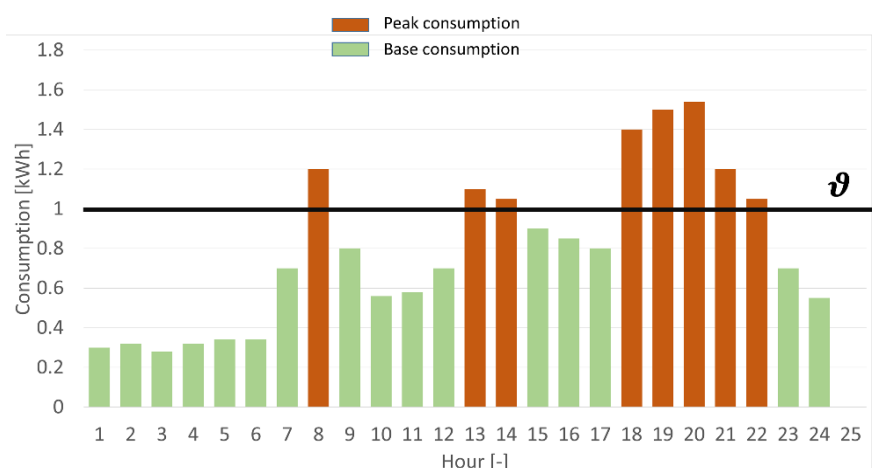

*Figure A10. Typical IPP grid tariff design applied to an individual household's consumption pattern.*

Hours that are within the top percentages defined above the peak threshold are defined in $H^{peak,\theta}$ and all consumption in those hours is defined as peak consumption. The residual hours that are in the lower range of the load duration curve within $H^{base,\theta}$ are defined as base consumption. Figure A11 gives in Appendix A.1. visualizes how peak and base hours are defined and grid tariffs apply.
Figure A11 shows how DCPP categorizes consumption into base and peak.

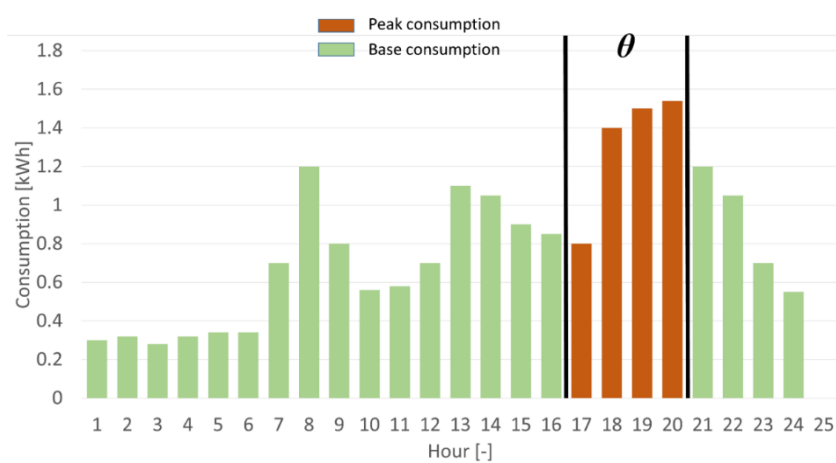

*Figure A11. Typical DCPP grid tariff design applied to individual household consumption.*

Assuming that the timeframe from 5 pm to 9 pm are within the top $\theta$ =1% of the national load duration curve. All consumption, irrespective of the level categorizes as peak consumption within those hours (red bars). Consumption in the residual 99% of the load duration curve qualifies as base consumption (green bars).

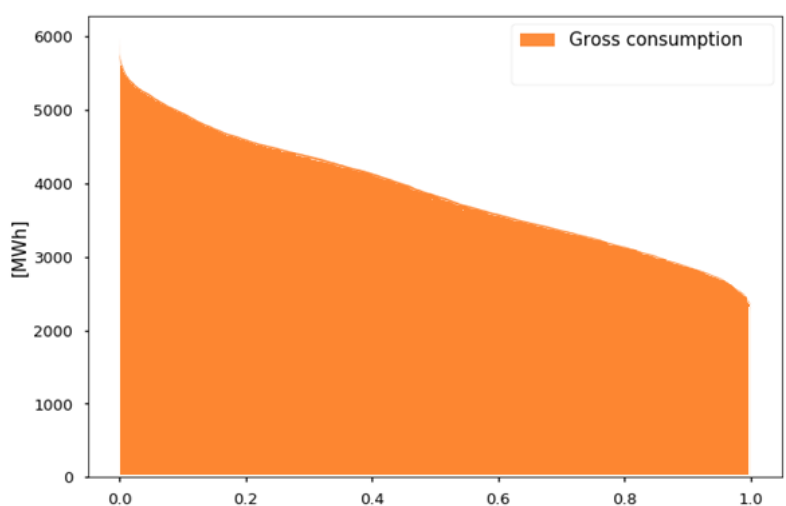

*Figure A12 Load-duration curve of Denmark in 2017. The shape is comparable to other industrialized countries. Denmark has in contrast to other industrialized countries a slightly higher share of residential demand and thus has a little bit less baseload as compared to e.g. Germany with more industrial baseload.*

**Danish context in grid tariff design**

    European countries are undergoing a rapid increase in heat and transport electrification driven by ambitious energy policies and rising imported energy prices (European Parliament Legislative Observatory, 2022). The take-off of electric cars and heat pumps is especially visible in countries also benefiting from low electricity prices, like Denmark, where wind and solar energy meet half of the

electricity consumption (Energistyrelsen, 2021, n.d.; Eurostat, 2018). However, this upcoming demand questions the room for maneuver that existing distribution grid oversizing grants them (Hansen et al., 2021). Subsequently, the question arises of which consumers and consumption types should pay more or less for accessing a grid that was not initially sized to meet new electricity and power needs.

With the recent announcement from the E.U. Commission mandating utilities to define their cost recovery methods to accelerate household electrification, European utilities are now urged to develop new tariff setups capable of transparently sending appropriate signals that smooth out peaks (Copenhagen Economics and VVA Europe, 2016). Rather than radical tariff transformation, updates use incremental steps to allow consumers to adapt to changes over time (Dansk Energi, 2022). Smaller steps help customers adjust to the new signals and avoid certain groups suddenly being exposed to significant cost increases from one year to another. In this context, Danish and European system operators are interested in understanding the dynamics and the magnitude of changes in different customer groups.

### A.2. Additional results

*Table A4. New grid tariffs for base and peak consumption per scenario from a flat volumetric tariff of 0.025 €/kWh (multiplied by a recovery factor of 0.95).*

| Scenario | Base grid tariff [€/kWh] | Peak grid tariff [€/kWh] |
|---|---|---|
| IPP; 1kWh | 0.023 | 0.027 |
| IPP; 1.5kWh | 0.023 | 0.031 |
| IPP; 2kWh | 0.023 | 0.036 |
| IPP; 3kWh | 0.023 | 0.071 |
| DCPP; 1% | 0.023 | 0.094 |
| DCPP; 5% | 0.023 | 0.041 |
| DCPP; 20% | 0.023 | 0.028 |
| DCPP; 40% | 0.023 | 0.026 |
| DCIPP; (2kWh, 1%) | 0.023 | 0.368 |
| DCIPP; (2kWh, 5%) | 0.023 | 0.131 |
| DCIPP; (2kWh, 20%) | 0.023 | 0.058 |
| DCIPP; (2kWh, 40%) | 0.023 | 0.044 |

**Impact of IPP on dwelling types and sizes**

Figure A13 shows the average relative cost redistribution compared to flat volumetric tariffs for households that do not own an EV or HP and that live in houses and apartments of different sizes. The average is calculated across all occupancy and income characteristics. Figure 4 shows the relative redistribution of grid cost with a technology focus. The figure focuses on a single-family house (H) in the large dwelling area (A3) and income group (€3) owning an EV, HP, or neither (No-tech) to isolate the impact on technology ownership. The investigated IPP ranges from a 1 *kWh/h* to a 3 *kWh/h* threshold.

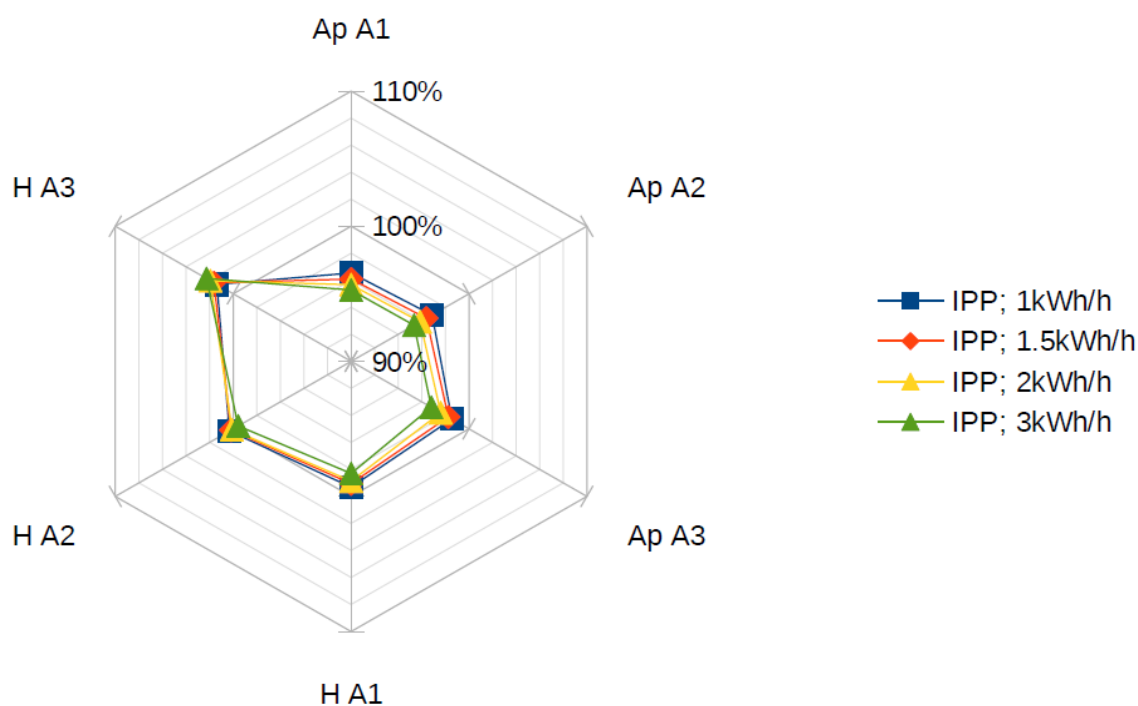

*Figure A13. Average relative redistribution of volumetric grid cost with IPP on Apartments (Ap) and Houses (H) with different dwelling areas (A1-A3). No EVs or HPs are present in the households.*

Figure A13 shows that most households would benefit from IPP. The smaller the dwelling size, the larger the reduction of volumetric grid costs compared to flat tariffs. Consequently, apartments could save from IPP because the living areas are on average smaller than detached houses. Only large detached houses (H A3) would face increasing costs, and only in some scenarios.

Moving the threshold to higher levels has two further impacts on households. First, the higher the threshold, the greater the annual savings for a small apartment ranging from 3.46% to 4.72%. At the same time, when the dwelling area increase, the savings are reduced. A large apartment has savings of between 1.45% and 3.19%. The same tendency is also observed in houses, though on a different level depending on the dwelling area. Small houses (A1) have decreasing costs only rarely, being almost unchanged between 0.68% and 1.7% with a 3 *kWh/h* threshold. Medium and large houses almost always pay more for IPP. A large house faces a rise of around 2.27% with a 3 *kWh/h* threshold.

**Impact of DCPP on dwelling types and sizes**

DCPP is applied at four different levels of national load, starting from the top 1%, 5%, 20% and 40%. Rather than the actual numbers, the shape of the radar plots is more important in showing the redistribution across different categories. **Error! Reference source not found.** shows the average relative cost redistribution per dwelling type and area. Figure A14 zooms in on the redistribution of the DCPP for a household living in a single-family house in the large dwelling area and income group and owning either an EV, a HP or neither.

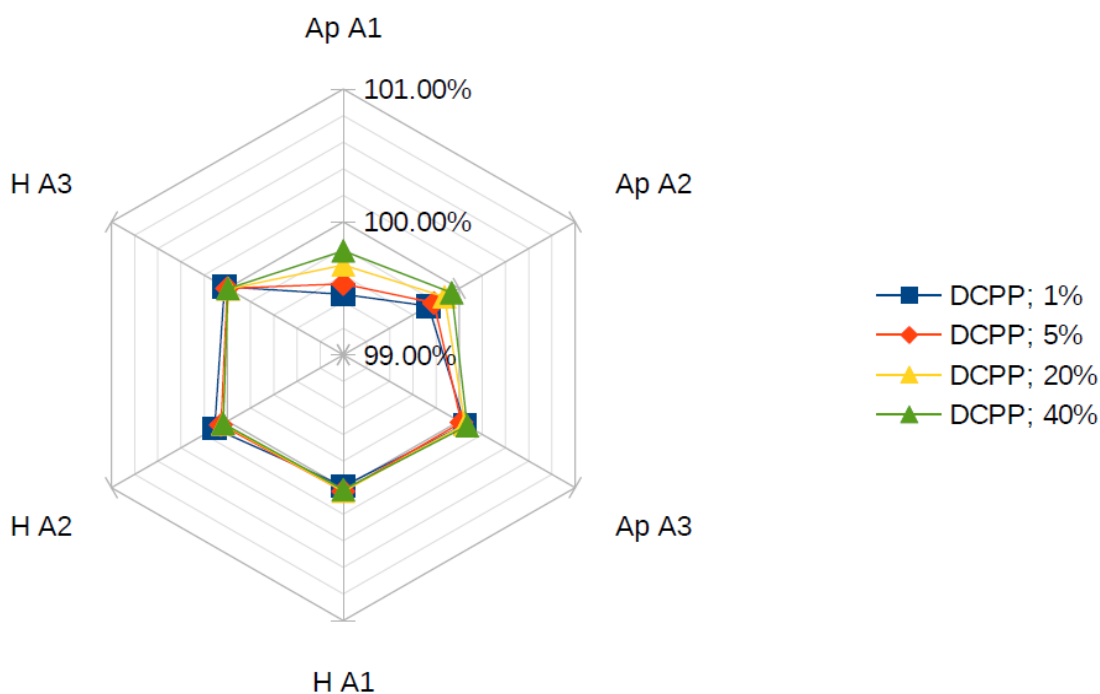

*Figure A14. Average relative redistribution volumetric grid cost with DCPP on Apartments (Ap) and Houses (H) with different dwelling areas (A1-A3). No EVs or HPs are present in the households.*

Although households in small apartments benefit the most from DCPP, as shown in Figure A14, this benefit shrinks the more hours are included in the DCPP (1% towards 40%), going from 0.54% to 0.22% in savings. The larger the dwelling size of the apartment, the smaller the savings. Large apartments pay slightly more, from 0.05% to 0.07%. Similarly, the larger the house, the higher the relative increase in grid costs. In contrast to the apartments, the more hours that are included in the DCPP, the fewer the large households have to pay in additional yearly network costs. Noticeably, a medium-sized house has to pay 0.11% more in the 1% DCPP scenario, while large houses are barely affected. The small changes are purely due to the chosen $f^{recov}$ of 0.95 as introduced in section 2.2. to maintain comparability with the IPP scenario. Differences increase with higher recovery factors and outcomes for that are in appendix Table A6-Table A9 and in the supplementary material.

*Table A5. Difference in total yearly cost paid per household on volumetric charges for each design in the respective household category in €/year compared to the base case of a flat volumetric design.*

| Category | IPP; 1kWh | IPP; 1.5kWh | IPP; 2kWh | IPP; 3kWh | DCPP; 1% | DCPP; 5% | DCPP; 20% | DCPP; 40% | DCIPP; (2kWh/h, 1%) | DCIPP; (2kWh/h, 5%) | DCIPP; (2kWh/h, 20%) | DCIPP; (2kWh/h, 40%) |
|---|---|---|---|---|---|---|---|---|---|---|---|---|
| Ap_P1_A1_€1_EV0_HP0 | -1.10 | -1.21 | -1.31 | -1.38 | -0.21 | -0.16 | -0.11 | -0.08 | -1.31 | -1.30 | -1.30 | -1.30 |
| Ap_P1_A1_€1_EV0_HP1 | 2.34 | 1.88 | 0.40 | -1.84 | 0.70 | 1.44 | 0.85 | 0.14 | 1.93 | 3.33 | 2.00 | 0.96 |
| Ap_P1_A1_€2_EV0_HP0 | -1.03 | -1.16 | -1.26 | -1.36 | -0.25 | -0.24 | -0.18 | -0.12 | -1.30 | -1.28 | -1.26 | -1.26 |
| Ap_P1_A1_€2_EV0_HP1 | 2.63 | 2.10 | 0.98 | 0.00 | 0.70 | 1.27 | 0.70 | 0.03 | 2.44 | 3.40 | 2.32 | 1.46 |
| Ap_P1_A1_€3_EV0_HP0 | -0.88 | -1.05 | -1.28 | -1.34 | -0.32 | -0.28 | -0.19 | -0.13 | -1.31 | -1.31 | -1.29 | -1.29 |

| Category | IPP; 1kWh | IPP; 1.5kWh | IPP; 2kWh | IPP; 3kWh | DCPP; 1% | DCPP; 5% | DCPP; 20% | DCPP; 40% | DCIPP; (2kWh/h, 1%) | DCIPP; (2kWh/h, 5%) | DCIPP; (2kWh/h, 20%) | DCIPP; (2kWh/h, 40%) |
|---|---|---|---|---|---|---|---|---|---|---|---|---|
| Ap_P1_A2_€1_EV0_HP0 | -1.17 | -1.30 | -1.41 | -1.51 | -0.15 | -0.09 | -0.05 | -0.03 | -1.42 | -1.41 | -1.40 | -1.40 |
| Ap_P1_A2_€1_EV0_HP1 | 3.76 | 4.15 | 2.46 | 0.55 | 0.97 | 1.83 | 1.14 | 0.30 | 5.54 | 7.85 | 5.54 | 3.79 |
| Ap_P1_A2_€2_EV0_HP0 | -1.09 | -1.23 | -1.39 | -1.54 | -0.22 | -0.20 | -0.15 | -0.10 | -1.41 | -1.40 | -1.38 | -1.38 |
| Ap_P1_A2_€2_EV0_HP1 | 2.64 | 1.67 | 0.30 | -1.07 | 0.91 | 1.60 | 0.86 | 0.12 | 1.99 | 2.65 | 1.45 | 0.85 |
| Ap_P1_A2_€3_EV0_HP0 | -1.03 | -1.18 | -1.34 | -1.54 | -0.32 | -0.25 | -0.17 | -0.12 | -1.44 | -1.42 | -1.39 | -1.37 |
| Ap_P1_A3_€1_EV0_HP0 | -1.23 | -1.37 | -1.52 | -1.64 | -0.12 | -0.04 | -0.02 | -0.01 | -1.57 | -1.53 | -1.51 | -1.51 |
| Ap_P1_A3_€1_EV0_HP1 | 5.59 | 8.46 | 9.30 | 11.46 | 1.14 | 2.12 | 1.32 | 0.35 | 10.91 | 15.91 | 12.52 | 9.89 |
| Ap_P1_A3_€2_EV0_HP0 | -1.12 | -1.29 | -1.49 | -1.66 | -0.22 | -0.17 | -0.12 | -0.08 | -1.58 | -1.54 | -1.51 | -1.50 |
| Ap_P1_A3_€2_EV0_HP1 | 6.51 | 9.70 | 11.52 | 19.40 | 1.52 | 2.25 | 1.21 | 0.18 | 13.74 | 17.57 | 13.61 | 11.23 |
| Ap_P1_A3_€3_EV0_HP0 | -1.07 | -1.30 | -1.58 | -1.95 | -0.36 | -0.25 | -0.15 | -0.11 | -1.71 | -1.67 | -1.60 | -1.58 |
| Ap_P2_A1_€1_EV0_HP0 | -1.03 | -1.23 | -1.45 | -1.63 | -0.05 | -0.06 | -0.02 | 0.00 | -1.45 | -1.44 | -1.44 | -1.44 |
| Ap_P2_A1_€1_EV0_HP1 | 4.30 | 4.68 | 3.02 | 1.30 | 1.34 | 1.87 | 1.08 | 0.29 | 6.97 | 7.84 | 5.64 | 4.22 |
| Ap_P2_A1_€2_EV0_HP0 | -0.95 | -1.17 | -1.44 | -1.73 | 0.06 | 0.00 | 0.01 | 0.03 | -1.37 | -1.39 | -1.41 | -1.41 |
| Ap_P2_A1_€2_EV0_HP1 | 4.07 | 5.06 | 4.81 | 5.62 | 1.02 | 1.57 | 0.96 | 0.28 | 7.30 | 8.80 | 7.41 | 6.05 |
| Ap_P2_A1_€3_EV0_HP0 | -0.93 | -1.16 | -1.42 | -1.67 | 0.01 | -0.04 | -0.03 | 0.00 | -1.41 | -1.41 | -1.41 | -1.40 |
| Ap_P2_A2_€1_EV0_HP0 | -1.06 | -1.29 | -1.58 | -1.83 | 0.00 | -0.04 | -0.01 | 0.02 | -1.54 | -1.58 | -1.57 | -1.56 |
| Ap_P2_A2_€1_EV0_HP1 | 6.74 | 9.31 | 10.12 | 12.07 | 1.23 | 1.82 | 1.07 | 0.26 | 12.30 | 14.09 | 12.51 | 10.87 |
| Ap_P2_A2_€2_EV0_HP0 | -1.03 | -1.30 | -1.64 | -1.99 | 0.12 | 0.04 | 0.05 | 0.06 | -1.56 | -1.61 | -1.60 | -1.60 |
| Ap_P2_A2_€2_EV0_HP1 | 5.28 | 6.68 | 6.23 | 5.05 | 1.22 | 1.92 | 1.19 | 0.43 | 9.91 | 12.47 | 9.91 | 8.10 |
| Ap_P2_A2_€3_EV0_HP0 | -0.96 | -1.23 | -1.58 | -1.94 | 0.02 | -0.04 | -0.02 | 0.01 | -1.49 | -1.56 | -1.54 | -1.54 |
| Ap_P2_A3_€1_EV0_HP0 | -0.91 | -1.10 | -1.42 | -1.76 | 0.00 | -0.01 | 0.03 | 0.04 | -1.47 | -1.42 | -1.38 | -1.39 |
| Ap_P2_A3_€1_EV0_HP1 | 5.70 | 9.79 | 12.90 | 23.88 | 0.86 | 1.47 | 0.85 | 0.22 | 11.60 | 17.01 | 14.97 | 12.53 |
| Ap_P2_A3_€2_EV0_HP0 | -0.94 | -1.24 | -1.65 | -2.10 | 0.14 | 0.13 | 0.13 | 0.11 | -1.62 | -1.59 | -1.57 | -1.58 |
| Ap_P2_A3_€2_EV0_HP1 | 8.11 | 13.23 | 15.99 | 18.66 | 1.66 | 2.68 | 1.71 | 0.63 | 19.19 | 26.94 | 22.48 | 18.45 |
| Ap_P2_A3_€3_EV0_HP0 | -0.56 | -0.77 | -1.25 | -1.82 | 0.08 | 0.10 | 0.09 | 0.07 | -1.21 | -1.16 | -1.13 | -1.15 |
| Ap_P2_A3_€3_EV0_HP1 | 8.90 | 14.83 | 18.96 | 26.98 | 1.38 | 2.21 | 1.26 | 0.32 | 20.09 | 26.69 | 22.87 | 19.79 |
| Ap_P3_A3_€3_EV0_HP0 | 0.36 | 0.44 | -0.10 | -1.03 | 0.33 | 0.06 | 0.06 | 0.10 | 0.32 | -0.09 | -0.06 | 0.01 |
| H_P1_A1_€1_EV0_HP0 | -1.21 | -1.43 | -1.52 | -1.67 | -0.17 | -0.03 | -0.02 | -0.04 | -1.62 | -1.50 | -1.51 | -1.54 |
| H_P1_A1_€1_EV0_HP1 | 6.49 | 8.73 | 8.76 | 9.83 | 0.79 | 2.00 | 1.11 | 0.10 | 10.24 | 15.44 | 11.47 | 8.71 |
| H_P1_A1_€2_EV0_HP0 | -0.98 | -1.20 | -1.31 | -1.59 | -0.30 | -0.21 | -0.17 | -0.15 | -1.55 | -1.42 | -1.39 | -1.39 |
| H_P1_A1_€2_EV0_HP1 | 8.49 | 11.66 | 11.51 | 12.39 | 0.52 | 1.63 | 0.83 | -0.12 | 11.08 | 15.75 | 12.40 | 10.17 |
| H_P1_A1_€3_EV0_HP0 | -0.90 | -0.97 | -1.14 | -1.29 | -0.48 | -0.33 | -0.24 | -0.20 | -1.52 | -1.28 | -1.20 | -1.20 |
| H_P1_A2_€1_EV0_HP0 | -1.12 | -1.32 | -1.35 | -1.32 | -0.19 | 0.00 | 0.01 | -0.05 | -1.60 | -1.31 | -1.33 | -1.40 |
| H_P1_A2_€1_EV0_HP1 | 10.33 | 15.99 | 17.21 | 19.01 | 1.10 | 2.37 | 1.39 | 0.18 | 19.35 | 26.90 | 21.60 | 17.45 |
| H_P1_A2_€2_EV0_HP0 | -0.99 | -1.19 | -1.21 | -1.15 | -0.31 | -0.20 | -0.16 | -0.16 | -1.59 | -1.41 | -1.35 | -1.34 |
| H_P1_A2_€2_EV0_HP1 | 10.22 | 15.42 | 16.93 | 18.12 | 0.97 | 2.09 | 1.01 | -0.10 | 18.49 | 25.04 | 19.17 | 15.77 |
| H_P1_A2_€3_EV0_HP0 | -0.82 | -0.90 | -1.15 | -1.36 | -0.52 | -0.32 | -0.24 | -0.21 | -1.64 | -1.29 | -1.23 | -1.25 |
| H_P1_A3_€1_EV0_HP0 | -0.63 | -0.42 | -0.36 | 0.62 | -0.27 | -0.01 | -0.02 | -0.08 | -1.01 | -0.40 | -0.44 | -0.52 |
| H_P1_A3_€1_EV0_HP1 | 14.35 | 23.36 | 28.96 | 42.74 | 1.17 | 2.70 | 1.40 | 0.00 | 27.18 | 38.99 | 31.90 | 26.66 |
| H_P1_A3_€2_EV0_HP0 | -0.71 | -0.63 | -0.77 | -0.40 | -0.38 | -0.19 | -0.17 | -0.18 | -1.34 | -1.00 | -0.98 | -0.98 |

| Category | IPP; 1kWh | IPP; 1.5kWh | IPP; 2kWh | IPP; 3kWh | DCPP; 1% | DCPP; 5% | DCPP; 20% | DCPP; 40% | DCIPP; (2kWh/h, 1%) | DCIPP; (2kWh/h, 5%) | DCIPP; (2kWh/h, 20%) | DCIPP; (2kWh/h, 40%) |
|---|---|---|---|---|---|---|---|---|---|---|---|---|
| H_P1_A3_€2_EV0_HP1 | 14.63 | 25.73 | 35.88 | 62.78 | 0.96 | 2.41 | 1.15 | -0.19 | 29.36 | 42.60 | 36.62 | 31.28 |
| H_P1_A3_€3_EV0_HP0 | -0.10 | 0.23 | 0.35 | 1.67 | -0.56 | -0.25 | -0.21 | -0.24 | -0.74 | 0.04 | 0.08 | 0.03 |
| H_P2_A1_€1_EV0_HP0 | -0.72 | -0.87 | -1.25 | -1.63 | 0.03 | 0.11 | 0.10 | 0.06 | -1.30 | -1.10 | -1.13 | -1.19 |
| H_P2_A1_€2_EV0_HP0 | -0.60 | -1.04 | -1.31 | -1.89 | 0.16 | 0.21 | 0.16 | 0.11 | -1.22 | -1.07 | -1.11 | -1.18 |
| H_P2_A1_€2_EV0_HP1 | 9.92 | 14.75 | 16.43 | 20.93 | 1.32 | 2.46 | 1.35 | 0.29 | 20.31 | 27.36 | 21.54 | 17.60 |
| H_P2_A1_€3_EV0_HP0 | -0.12 | -0.47 | -0.69 | -1.38 | 0.00 | 0.00 | 0.00 | 0.00 | -0.74 | -0.67 | -0.62 | -0.62 |
| H_P2_A1_€3_EV0_HP1 | 13.42 | 21.71 | 26.88 | 34.27 | 1.08 | 2.38 | 1.29 | 0.11 | 26.24 | 36.44 | 31.14 | 26.78 |
| H_P2_A2_€1_EV0_HP0 | -0.27 | -0.26 | -0.60 | -0.86 | -0.07 | 0.09 | 0.08 | 0.02 | -0.82 | -0.42 | -0.44 | -0.54 |
| H_P2_A2_€1_EV0_HP1 | 13.86 | 21.32 | 23.35 | 24.10 | 1.25 | 2.67 | 1.44 | 0.17 | 27.20 | 37.26 | 29.49 | 24.23 |
| H_P2_A2_€2_EV0_HP0 | -0.52 | -1.03 | -1.33 | -1.89 | 0.20 | 0.30 | 0.22 | 0.14 | -1.16 | -0.95 | -1.04 | -1.15 |
| H_P2_A2_€2_EV0_HP1 | 12.43 | 19.37 | 22.51 | 28.94 | 1.21 | 2.60 | 1.51 | 0.33 | 24.83 | 35.17 | 29.01 | 23.91 |
| H_P2_A2_€3_EV0_HP0 | -0.04 | -0.42 | -0.66 | -1.35 | 0.02 | 0.01 | 0.00 | 0.00 | -0.68 | -0.68 | -0.59 | -0.58 |
| H_P2_A2_€3_EV0_HP1 | 14.71 | 23.93 | 29.45 | 39.10 | 1.11 | 2.27 | 1.17 | 0.06 | 29.72 | 39.19 | 33.50 | 29.28 |
| H_P2_A3_€2_EV0_HP0 | -0.01 | -0.39 | -0.53 | -0.69 | 0.15 | 0.31 | 0.23 | 0.13 | -0.46 | -0.07 | -0.19 | -0.34 |
| H_P2_A3_€2_EV0_HP1 | 17.85 | 30.48 | 40.60 | 59.97 | 1.35 | 2.98 | 1.62 | 0.18 | 37.34 | 53.38 | 46.01 | 39.20 |
| H_P2_A3_€3_EV0_HP0 | 0.72 | 0.60 | 0.65 | 0.78 | -0.09 | 0.04 | 0.01 | -0.02 | 0.28 | 0.66 | 0.72 | 0.69 |
| H_P2_A3_€3_EV0_HP1 | 19.48 | 34.25 | 46.75 | 76.83 | 1.29 | 2.82 | 1.36 | -0.06 | 40.40 | 56.40 | 49.33 | 43.35 |
| H_P3_A1_€1_EV0_HP0 | 0.50 | 0.86 | 0.65 | 0.66 | -0.02 | -0.02 | 0.02 | 0.03 | 0.41 | 0.52 | 0.65 | 0.66 |
| H_P3_A1_€2_EV0_HP0 | 0.69 | 0.67 | 0.70 | 0.02 | 0.29 | 0.10 | 0.09 | 0.12 | 1.07 | 0.70 | 0.70 | 0.79 |
| H_P3_A1_€3_EV0_HP0 | 1.27 | 1.37 | 1.48 | 0.80 | 0.31 | 0.03 | 0.01 | 0.06 | 1.97 | 1.34 | 1.38 | 1.52 |
| H_P3_A1_€3_EV0_HP1 | 17.92 | 30.08 | 38.58 | 58.67 | 1.50 | 2.48 | 1.37 | 0.23 | 37.71 | 47.49 | 42.58 | 38.69 |
| H_P3_A2_€1_EV0_HP0 | 1.33 | 2.08 | 2.07 | 2.62 | -0.09 | 0.00 | 0.03 | 0.01 | 1.66 | 2.07 | 2.18 | 2.15 |
| H_P3_A2_€2_EV0_HP0 | 1.17 | 1.29 | 1.47 | 1.04 | 0.22 | 0.07 | 0.07 | 0.09 | 1.77 | 1.46 | 1.45 | 1.54 |
| H_P3_A2_€2_EV0_HP1 | 12.94 | 20.88 | 25.88 | 39.04 | 1.06 | 2.01 | 1.09 | 0.18 | 25.17 | 32.59 | 29.52 | 26.41 |
| H_P3_A2_€3_EV0_HP0 | 1.44 | 1.52 | 1.61 | 0.80 | 0.29 | -0.02 | -0.02 | 0.04 | 2.21 | 1.44 | 1.51 | 1.66 |
| H_P3_A2_€3_EV0_HP1 | 14.60 | 24.79 | 32.43 | 49.44 | 1.23 | 1.83 | 0.89 | 0.05 | 30.95 | 37.90 | 33.94 | 31.36 |
| H_P3_A2_€3_EV1_HP0 | 6.64 | 13.38 | 22.86 | 76.25 | -0.05 | -0.58 | -0.65 | -0.55 | 10.80 | 10.41 | 11.68 | 13.67 |
| H_P3_A3_€1_EV0_HP0 | 2.73 | 4.05 | 4.40 | 6.34 | -0.22 | -0.06 | -0.04 | -0.06 | 3.49 | 3.97 | 4.15 | 4.19 |
| H_P3_A3_€2_EV0_HP0 | 2.38 | 3.06 | 3.89 | 5.09 | 0.05 | 0.01 | 0.01 | 0.02 | 3.55 | 3.58 | 3.65 | 3.80 |
| H_P3_A3_€2_EV0_HP1 | 18.95 | 33.49 | 46.95 | 86.21 | 1.24 | 2.48 | 1.25 | 0.07 | 39.15 | 53.02 | 48.21 | 43.82 |
| H_P3_A3_€3_EV0_HP0 | 2.32 | 2.71 | 3.13 | 3.07 | 0.11 | -0.11 | -0.09 | -0.03 | 3.40 | 2.78 | 2.92 | 3.08 |
| H_P3_A3_€3_EV0_HP1 | 19.02 | 33.11 | 45.55 | 78.76 | 1.17 | 2.20 | 1.00 | -0.08 | 38.50 | 50.07 | 45.52 | 42.12 |
| H_P3_A3_€3_EV1_HP1 | 13.68 | 27.30 | 47.04 | 159.06 | -1.30 | -1.84 | -1.67 | -1.36 | 17.29 | 17.36 | 21.57 | 26.69 |
| H_P5+_A1_€1_EV0_HP0 | 2.74 | 4.48 | 5.50 | 8.77 | -0.14 | -0.08 | 0.02 | 0.02 | 3.80 | 4.36 | 4.72 | 5.01 |
| H_P5+_A1_€2_EV0_HP0 | 2.06 | 3.15 | 3.30 | 3.21 | 0.11 | 0.00 | 0.09 | 0.16 | 3.02 | 2.66 | 2.97 | 3.34 |
| H_P5+_A1_€3_EV0_HP0 | 2.97 | 4.46 | 4.75 | 5.50 | 0.43 | 0.12 | 0.12 | 0.14 | 5.29 | 4.52 | 4.63 | 4.82 |
| H_P5+_A2_€1_EV0_HP0 | 2.99 | 4.68 | 5.30 | 6.69 | -0.22 | -0.10 | 0.00 | 0.02 | 3.82 | 4.42 | 4.81 | 5.07 |
| H_P5+_A2_€2_EV0_HP0 | 2.81 | 4.45 | 5.09 | 6.73 | 0.09 | -0.01 | 0.07 | 0.14 | 4.64 | 4.40 | 4.65 | 5.09 |
| H_P5+_A2_€3_EV0_HP0 | 2.73 | 3.36 | 3.96 | 3.73 | 0.43 | 0.07 | 0.07 | 0.14 | 4.94 | 3.76 | 3.83 | 4.08 |
| H_P5+_A2_€3_EV0_HP1 | 17.58 | 30.97 | 42.04 | 74.95 | 1.25 | 1.99 | 1.12 | 0.25 | 36.37 | 45.60 | 42.93 | 40.62 |

| Category | IPP; 1kWh | IPP; 1.5kWh | IPP; 2kWh | IPP; 3kWh | DCPP; 1% | DCPP; 5% | DCPP; 20% | DCPP; 40% | DCIPP; (2kWh/h, 1%) | DCIPP; (2kWh/h, 5%) | DCIPP; (2kWh/h, 20%) | DCIPP; (2kWh/h, 40%) |
|---|---|---|---|---|---|---|---|---|---|---|---|---|
| H_P5+_A3_€1_EV0_HP0 | 6.42 | 10.14 | 12.58 | 20.57 | -0.04 | 0.11 | 0.12 | 0.06 | 10.55 | 11.71 | 12.20 | 12.41 |
| H_P5+_A3_€2_EV0_HP0 | 4.87 | 7.57 | 9.00 | 12.70 | 0.07 | -0.01 | 0.05 | 0.09 | 8.04 | 7.74 | 8.13 | 8.74 |
| H_P5+_A3_€3_EV0_HP0 | 3.89 | 5.03 | 6.18 | 7.37 | 0.31 | -0.02 | -0.01 | 0.07 | 6.78 | 5.68 | 5.82 | 6.13 |
| H_P5+_A3_€3_EV0_HP1 | 21.33 | 37.41 | 51.59 | 89.16 | 1.58 | 2.47 | 1.17 | 0.08 | 45.57 | 57.40 | 52.64 | 49.19 |
| H_P5+_A3_€3_EV1_HP0 | 15.31 | 30.25 | 52.20 | 180.00 | -1.39 | -1.86 | -1.64 | -1.34 | 20.28 | 20.75 | 25.10 | 30.25 |

*Table A6. Sensitivity study on $f^{red}$ in the IPP scenario with a threshold of 1 kWh/h. The numbers show the relative change to the flat volumetric tariff.*

| Dwelling type | Dwelling area | $f^{red}$=0.95 | $f^{red}$=0.9 | $f^{red}$=0.8 |
|---|---|---|---|---|
| Ap | A1 | 0.9654 | 0.9308 | 0.8615 |
| Ap | A2 | 0.9686 | 0.9373 | 0.8745 |
| Ap | A3 | 0.9855 | 0.9709 | 0.9419 |
| H | A1 | 0.9932 | 0.9864 | 0.9728 |
| H | A2 | 1.0032 | 1.0065 | 1.0129 |
| H | A3 | 1.0139 | 1.0277 | 1.0554 |

*Table A7. Sensitivity study on $f^{red}$ in the DCPP scenario with a trigger percentage of 1%. The numbers show the relative change to the flat volumetric tariff.*

| Dwelling type | Dwelling area | $f^{red}$=0.95 | $f^{red}$=0.9 | $f^{red}$=0.8 |
|---|---|---|---|---|
| Ap | A1 | 0.9946 | 0.9891 | 0.9782 |
| Ap | A2 | 0.9973 | 0.9947 | 0.9894 |
| Ap | A3 | 1.0005 | 1.0009 | 1.0019 |
| H | A1 | 0.9999 | 0.9998 | 0.9996 |
| H | A2 | 1.0011 | 1.0022 | 1.0044 |
| H | A3 | 1.0002 | 1.0005 | 1.0010 |

*Table A8. Sensitivity study on $f^{red}$ in the DCIPP scenario with a threshold of 1 kWh/h and a trigger percentage of 1%. The numbers show the relative change to the flat volumetric tariff.*

| Dwelling type | Dwelling area | $f^{red}$=0.95 | $f^{red}$=0.9 | $f^{red}$=0.8 |
|---|---|---|---|---|
| Ap | A1 | 0.9675 | 0.9350 | 0.8701 |
| Ap | A2 | 0.9718 | 0.9436 | 0.8872 |
| Ap | A3 | 0.9887 | 0.9773 | 0.9546 |
| H | A1 | 0.9945 | 0.9889 | 0.9778 |
| H | A2 | 1.0049 | 1.0099 | 1.0198 |
| H | A3 | 1.0112 | 1.0224 | 1.0447 |

*Table A9. Sensitivity on $f^{red}$ and the resulting base and peak prices per scenario.*

| Scenario | f^{red}=0.95 | f^{red}=0.9 | f^{red}=0.8 |
|---|---|---|---|
| Base | 17.34 | 16.43 | 14.60 |
| IPP 1kWh | 20.29 | 22.34 | 26.43 |
| DCPP 1% | 70.05 | 121.84 | 225.44 |
| DCIPP (1kWh, 1%) | 116.08 | 213.92 | 409.58 |

Table A10 Savings potential per household category and grid tariff design. Percentage of total cost saving for shifting 1% of peak consumption to base consumption.

| Category | IPP; 1kWh | IPP; 1.5kWh | IPP; 2kWh | IPP; 3kWh | DCPP; 1% | DCPP; 5% | DCPP; 20% | DCPP; 40% | DCIPP; (2kWh/h, 1%) | DCIPP; (2kWh/h, 5%) | DCIPP; (2kWh/h, 20%) | DCIPP; (2kWh/h, 40%) |
|---|---|---|---|---|---|---|---|---|---|---|---|---|
| Ap_P1_A1_€1_EV0_HP0 | 0.012% | 0.009% | 0.005% | 0.002% | 0.043% | 0.045% | 0.046% | 0.047% | 0.005% | 0.005% | 0.005% | 0.005% |
| Ap_P1_A1_€1_EV0_HP1 | 0.071% | 0.067% | 0.054% | 0.033% | 0.056% | 0.063% | 0.058% | 0.051% | 0.067% | 0.079% | 0.068% | 0.059% |
| Ap_P1_A1_€2_EV0_HP0 | 0.015% | 0.010% | 0.007% | 0.003% | 0.042% | 0.042% | 0.044% | 0.046% | 0.005% | 0.006% | 0.006% | 0.006% |
| Ap_P1_A1_€2_EV0_HP1 | 0.073% | 0.068% | 0.059% | 0.050% | 0.056% | 0.061% | 0.056% | 0.050% | 0.071% | 0.079% | 0.070% | 0.063% |
| Ap_P1_A1_€3_EV0_HP0 | 0.020% | 0.014% | 0.005% | 0.003% | 0.039% | 0.041% | 0.044% | 0.046% | 0.004% | 0.004% | 0.005% | 0.005% |
| Ap_P1_A2_€1_EV0_HP0 | 0.013% | 0.009% | 0.006% | 0.002% | 0.045% | 0.047% | 0.048% | 0.049% | 0.005% | 0.006% | 0.006% | 0.006% |
| Ap_P1_A2_€1_EV0_HP1 | 0.079% | 0.082% | 0.069% | 0.054% | 0.058% | 0.064% | 0.059% | 0.052% | 0.092% | 0.109% | 0.092% | 0.079% |
| Ap_P1_A2_€2_EV0_HP0 | 0.017% | 0.013% | 0.008% | 0.003% | 0.044% | 0.044% | 0.046% | 0.047% | 0.007% | 0.007% | 0.008% | 0.008% |
| Ap_P1_A2_€2_EV0_HP1 | 0.072% | 0.064% | 0.053% | 0.041% | 0.058% | 0.064% | 0.057% | 0.051% | 0.067% | 0.073% | 0.062% | 0.057% |
| Ap_P1_A2_€3_EV0_HP0 | 0.018% | 0.013% | 0.008% | 0.002% | 0.040% | 0.042% | 0.045% | 0.046% | 0.005% | 0.006% | 0.007% | 0.007% |
| Ap_P1_A3_€1_EV0_HP0 | 0.018% | 0.014% | 0.010% | 0.006% | 0.047% | 0.049% | 0.050% | 0.050% | 0.008% | 0.009% | 0.010% | 0.010% |
| Ap_P1_A3_€1_EV0_HP1 | 0.092% | 0.112% | 0.117% | 0.132% | 0.059% | 0.066% | 0.060% | 0.053% | 0.128% | 0.159% | 0.138% | 0.121% |
| Ap_P1_A3_€2_EV0_HP0 | 0.023% | 0.019% | 0.014% | 0.009% | 0.045% | 0.046% | 0.047% | 0.048% | 0.011% | 0.012% | 0.013% | 0.013% |
| Ap_P1_A3_€2_EV0_HP1 | 0.093% | 0.113% | 0.124% | 0.169% | 0.060% | 0.065% | 0.058% | 0.051% | 0.137% | 0.159% | 0.136% | 0.122% |
| Ap_P1_A3_€3_EV0_HP0 | 0.027% | 0.022% | 0.016% | 0.008% | 0.042% | 0.045% | 0.047% | 0.048% | 0.013% | 0.014% | 0.016% | 0.016% |
| Ap_P2_A1_€1_EV0_HP0 | 0.022% | 0.016% | 0.010% | 0.004% | 0.049% | 0.048% | 0.049% | 0.050% | 0.010% | 0.010% | 0.010% | 0.010% |
| Ap_P2_A1_€1_EV0_HP1 | 0.085% | 0.088% | 0.075% | 0.061% | 0.061% | 0.066% | 0.059% | 0.052% | 0.105% | 0.112% | 0.095% | 0.084% |
| Ap_P2_A1_€2_EV0_HP0 | 0.025% | 0.019% | 0.012% | 0.004% | 0.051% | 0.050% | 0.050% | 0.051% | 0.014% | 0.014% | 0.013% | 0.013% |
| Ap_P2_A1_€2_EV0_HP1 | 0.082% | 0.090% | 0.088% | 0.094% | 0.058% | 0.063% | 0.058% | 0.052% | 0.106% | 0.117% | 0.107% | 0.097% |
| Ap_P2_A1_€3_EV0_HP0 | 0.024% | 0.018% | 0.010% | 0.003% | 0.050% | 0.049% | 0.049% | 0.050% | 0.011% | 0.011% | 0.011% | 0.011% |
| Ap_P2_A2_€1_EV0_HP0 | 0.025% | 0.020% | 0.012% | 0.006% | 0.050% | 0.050% | 0.050% | 0.050% | 0.013% | 0.012% | 0.013% | 0.013% |
| Ap_P2_A2_€1_EV0_HP1 | 0.092% | 0.108% | 0.112% | 0.123% | 0.058% | 0.062% | 0.057% | 0.052% | 0.125% | 0.134% | 0.126% | 0.117% |
| Ap_P2_A2_€2_EV0_HP0 | 0.027% | 0.021% | 0.013% | 0.005% | 0.053% | 0.051% | 0.051% | 0.051% | 0.015% | 0.014% | 0.014% | 0.014% |
| Ap_P2_A2_€2_EV0_HP1 | 0.085% | 0.094% | 0.091% | 0.084% | 0.058% | 0.063% | 0.058% | 0.053% | 0.114% | 0.130% | 0.114% | 0.103% |
| Ap_P2_A2_€3_EV0_HP0 | 0.028% | 0.022% | 0.013% | 0.004% | 0.050% | 0.049% | 0.050% | 0.050% | 0.015% | 0.014% | 0.014% | 0.014% |
| Ap_P2_A3_€1_EV0_HP0 | 0.033% | 0.029% | 0.023% | 0.016% | 0.050% | 0.050% | 0.051% | 0.051% | 0.022% | 0.023% | 0.024% | 0.024% |
| Ap_P2_A3_€1_EV0_HP1 | 0.091% | 0.119% | 0.139% | 0.202% | 0.057% | 0.061% | 0.056% | 0.052% | 0.131% | 0.164% | 0.152% | 0.138% |
| Ap_P2_A3_€2_EV0_HP0 | 0.033% | 0.028% | 0.020% | 0.012% | 0.052% | 0.052% | 0.052% | 0.052% | 0.021% | 0.021% | 0.021% | 0.021% |
| Ap_P2_A3_€2_EV0_HP1 | 0.098% | 0.126% | 0.140% | 0.154% | 0.060% | 0.066% | 0.061% | 0.054% | 0.156% | 0.193% | 0.172% | 0.153% |
| Ap_P2_A3_€3_EV0_HP0 | 0.041% | 0.038% | 0.031% | 0.021% | 0.051% | 0.051% | 0.051% | 0.051% | 0.031% | 0.032% | 0.032% | 0.032% |

| Category | IPP; 1kWh | IPP; 1.5kWh | IPP; 2kWh | IPP; 3kWh | DCPP; 1% | DCPP; 5% | DCPP; 20% | DCPP; 40% | DCIPP; (2kWh/h, 1%) | DCIPP; (2kWh/h, 5%) | DCIPP; (2kWh/h, 20%) | DCIPP; (2kWh/h, 40%) |
|---|---|---|---|---|---|---|---|---|---|---|---|---|
| Ap_P2_A3_€3_EV0_HP1 | 0.099% | 0.129% | 0.149% | 0.185% | 0.058% | 0.063% | 0.057% | 0.052% | 0.154% | 0.184% | 0.167% | 0.153% |
| Ap_P3_A3_€3_EV0_HP0 | 0.054% | 0.055% | 0.049% | 0.037% | 0.054% | 0.051% | 0.051% | 0.051% | 0.054% | 0.049% | 0.049% | 0.050% |
| H_P1_A1_€1_EV0_HP0 | 0.024% | 0.019% | 0.017% | 0.013% | 0.046% | 0.049% | 0.050% | 0.049% | 0.014% | 0.017% | 0.017% | 0.016% |
| H_P1_A1_€1_EV0_HP1 | 0.089% | 0.102% | 0.102% | 0.108% | 0.055% | 0.062% | 0.057% | 0.051% | 0.111% | 0.138% | 0.117% | 0.102% |
| H_P1_A1_€2_EV0_HP0 | 0.031% | 0.027% | 0.025% | 0.019% | 0.044% | 0.046% | 0.047% | 0.047% | 0.020% | 0.023% | 0.023% | 0.023% |
| H_P1_A1_€2_EV0_HP1 | 0.097% | 0.113% | 0.112% | 0.117% | 0.053% | 0.059% | 0.055% | 0.049% | 0.110% | 0.133% | 0.117% | 0.105% |
| H_P1_A1_€3_EV0_HP0 | 0.034% | 0.033% | 0.029% | 0.027% | 0.041% | 0.044% | 0.046% | 0.046% | 0.022% | 0.027% | 0.028% | 0.028% |
| H_P1_A2_€1_EV0_HP0 | 0.031% | 0.028% | 0.027% | 0.028% | 0.047% | 0.050% | 0.050% | 0.049% | 0.023% | 0.028% | 0.027% | 0.026% |
| H_P1_A2_€1_EV0_HP1 | 0.102% | 0.129% | 0.134% | 0.142% | 0.056% | 0.063% | 0.057% | 0.051% | 0.144% | 0.175% | 0.153% | 0.135% |
| H_P1_A2_€2_EV0_HP0 | 0.034% | 0.031% | 0.030% | 0.031% | 0.045% | 0.047% | 0.047% | 0.048% | 0.024% | 0.027% | 0.028% | 0.028% |
| H_P1_A2_€2_EV0_HP1 | 0.102% | 0.127% | 0.133% | 0.139% | 0.055% | 0.061% | 0.055% | 0.049% | 0.140% | 0.168% | 0.143% | 0.128% |
| H_P1_A2_€3_EV0_HP0 | 0.038% | 0.037% | 0.033% | 0.030% | 0.042% | 0.045% | 0.046% | 0.047% | 0.025% | 0.031% | 0.032% | 0.031% |
| H_P1_A3_€1_EV0_HP0 | 0.041% | 0.044% | 0.045% | 0.058% | 0.046% | 0.050% | 0.050% | 0.049% | 0.036% | 0.045% | 0.044% | 0.043% |
| H_P1_A3_€1_EV0_HP1 | 0.112% | 0.147% | 0.167% | 0.213% | 0.055% | 0.062% | 0.056% | 0.050% | 0.161% | 0.201% | 0.177% | 0.159% |
| H_P1_A3_€2_EV0_HP0 | 0.041% | 0.042% | 0.040% | 0.045% | 0.045% | 0.047% | 0.048% | 0.048% | 0.032% | 0.037% | 0.037% | 0.037% |
| H_P1_A3_€2_EV0_HP1 | 0.112% | 0.153% | 0.188% | 0.268% | 0.054% | 0.061% | 0.055% | 0.049% | 0.166% | 0.210% | 0.191% | 0.173% |
| H_P1_A3_€3_EV0_HP0 | 0.049% | 0.053% | 0.054% | 0.068% | 0.044% | 0.047% | 0.048% | 0.047% | 0.042% | 0.050% | 0.051% | 0.050% |
| H_P2_A1_€1_EV0_HP0 | 0.039% | 0.037% | 0.031% | 0.025% | 0.050% | 0.052% | 0.051% | 0.051% | 0.030% | 0.033% | 0.033% | 0.032% |
| H_P2_A1_€2_EV0_HP0 | 0.042% | 0.036% | 0.032% | 0.024% | 0.052% | 0.053% | 0.052% | 0.051% | 0.033% | 0.035% | 0.035% | 0.034% |
| H_P2_A1_€2_EV0_HP1 | 0.099% | 0.121% | 0.129% | 0.148% | 0.057% | 0.063% | 0.057% | 0.051% | 0.146% | 0.174% | 0.151% | 0.134% |
| H_P2_A1_€3_EV0_HP0 | 0.049% | 0.044% | 0.042% | 0.033% | 0.050% | 0.050% | 0.050% | 0.050% | 0.041% | 0.042% | 0.042% | 0.042% |
| H_P2_A1_€3_EV0_HP1 | 0.107% | 0.140% | 0.158% | 0.184% | 0.055% | 0.061% | 0.056% | 0.050% | 0.156% | 0.191% | 0.173% | 0.158% |
| H_P2_A2_€1_EV0_HP0 | 0.047% | 0.047% | 0.043% | 0.040% | 0.049% | 0.051% | 0.051% | 0.050% | 0.040% | 0.045% | 0.045% | 0.044% |
| H_P2_A2_€1_EV0_HP1 | 0.109% | 0.138% | 0.145% | 0.148% | 0.056% | 0.062% | 0.056% | 0.051% | 0.159% | 0.193% | 0.167% | 0.148% |
| H_P2_A2_€2_EV0_HP0 | 0.044% | 0.038% | 0.034% | 0.027% | 0.052% | 0.054% | 0.053% | 0.052% | 0.036% | 0.039% | 0.037% | 0.036% |
| H_P2_A2_€2_EV0_HP1 | 0.106% | 0.134% | 0.146% | 0.170% | 0.056% | 0.062% | 0.057% | 0.052% | 0.155% | 0.192% | 0.171% | 0.152% |
| H_P2_A2_€3_EV0_HP0 | 0.050% | 0.045% | 0.043% | 0.035% | 0.050% | 0.050% | 0.050% | 0.050% | 0.042% | 0.042% | 0.043% | 0.044% |
| H_P2_A2_€3_EV0_HP1 | 0.110% | 0.144% | 0.164% | 0.195% | 0.055% | 0.060% | 0.055% | 0.050% | 0.164% | 0.195% | 0.177% | 0.163% |
| H_P2_A3_€2_EV0_HP0 | 0.050% | 0.046% | 0.044% | 0.043% | 0.052% | 0.053% | 0.052% | 0.051% | 0.045% | 0.049% | 0.048% | 0.046% |
| H_P2_A3_€2_EV0_HP1 | 0.115% | 0.156% | 0.187% | 0.239% | 0.055% | 0.062% | 0.056% | 0.051% | 0.177% | 0.222% | 0.202% | 0.183% |
| H_P2_A3_€3_EV0_HP0 | 0.057% | 0.056% | 0.056% | 0.057% | 0.049% | 0.050% | 0.050% | 0.050% | 0.053% | 0.056% | 0.057% | 0.056% |
| H_P2_A3_€3_EV0_HP1 | 0.118% | 0.163% | 0.198% | 0.271% | 0.055% | 0.060% | 0.055% | 0.050% | 0.181% | 0.223% | 0.205% | 0.189% |
| H_P3_A1_€1_EV0_HP0 | 0.056% | 0.060% | 0.057% | 0.057% | 0.050% | 0.050% | 0.050% | 0.050% | 0.055% | 0.056% | 0.057% | 0.058% |
| H_P3_A1_€2_EV0_HP0 | 0.058% | 0.058% | 0.058% | 0.050% | 0.053% | 0.051% | 0.051% | 0.051% | 0.062% | 0.058% | 0.058% | 0.059% |
| H_P3_A1_€3_EV0_HP0 | 0.062% | 0.063% | 0.064% | 0.058% | 0.053% | 0.050% | 0.050% | 0.051% | 0.069% | 0.063% | 0.063% | 0.065% |
| H_P3_A1_€3_EV0_HP1 | 0.116% | 0.156% | 0.182% | 0.237% | 0.056% | 0.060% | 0.055% | 0.051% | 0.179% | 0.207% | 0.193% | 0.182% |
| H_P3_A2_€1_EV0_HP0 | 0.062% | 0.069% | 0.069% | 0.074% | 0.049% | 0.050% | 0.050% | 0.050% | 0.065% | 0.069% | 0.070% | 0.070% |
| H_P3_A2_€2_EV0_HP0 | 0.061% | 0.063% | 0.064% | 0.060% | 0.052% | 0.051% | 0.051% | 0.051% | 0.067% | 0.064% | 0.064% | 0.065% |
| H_P3_A2_€2_EV0_HP1 | 0.107% | 0.139% | 0.158% | 0.204% | 0.055% | 0.059% | 0.055% | 0.051% | 0.156% | 0.182% | 0.171% | 0.160% |
| H_P3_A2_€3_EV0_HP0 | 0.063% | 0.064% | 0.065% | 0.057% | 0.053% | 0.050% | 0.050% | 0.050% | 0.070% | 0.063% | 0.064% | 0.065% |

| Category | IPP; 1kWh | IPP; 1.5kWh | IPP; 2kWh | IPP; 3kWh | DCPP; 1% | DCPP; 5% | DCPP; 20% | DCPP; 40% | DCIPP; (2kWh/h, 1%) | DCIPP; (2kWh/h, 5%) | DCIPP; (2kWh/h, 20%) | DCIPP; (2kWh/h, 40%) |
|---|---|---|---|---|---|---|---|---|---|---|---|---|
| H_P3_A2_€3_EV0_HP1 | 0.110% | 0.148% | 0.174% | 0.227% | 0.055% | 0.058% | 0.054% | 0.050% | 0.169% | 0.192% | 0.179% | 0.171% |
| H_P3_A2_€3_EV1_HP0 | 0.091% | 0.129% | 0.178% | 0.374% | 0.050% | 0.046% | 0.046% | 0.046% | 0.115% | 0.113% | 0.120% | 0.131% |
| H_P3_A3_€1_EV0_HP0 | 0.071% | 0.080% | 0.083% | 0.096% | 0.048% | 0.050% | 0.050% | 0.050% | 0.076% | 0.080% | 0.081% | 0.081% |
| H_P3_A3_€2_EV0_HP0 | 0.069% | 0.074% | 0.081% | 0.090% | 0.050% | 0.050% | 0.050% | 0.050% | 0.078% | 0.078% | 0.079% | 0.080% |
| H_P3_A3_€2_EV0_HP1 | 0.117% | 0.162% | 0.200% | 0.293% | 0.055% | 0.059% | 0.055% | 0.050% | 0.178% | 0.216% | 0.203% | 0.191% |
| H_P3_A3_€3_EV0_HP0 | 0.068% | 0.071% | 0.075% | 0.074% | 0.051% | 0.049% | 0.049% | 0.050% | 0.077% | 0.072% | 0.073% | 0.074% |
| H_P3_A3_€3_EV0_HP1 | 0.117% | 0.161% | 0.196% | 0.277% | 0.054% | 0.058% | 0.054% | 0.050% | 0.176% | 0.208% | 0.196% | 0.186% |
| H_P3_A3_€3_EV1_HP0 | 0.106% | 0.155% | 0.218% | 0.450% | 0.044% | 0.042% | 0.043% | 0.044% | 0.120% | 0.120% | 0.135% | 0.153% |
| H_P5+_A1_€1_EV0_HP0 | 0.074% | 0.088% | 0.096% | 0.122% | 0.049% | 0.049% | 0.050% | 0.050% | 0.082% | 0.087% | 0.090% | 0.092% |
| H_P5+_A1_€2_EV0_HP0 | 0.069% | 0.078% | 0.079% | 0.079% | 0.051% | 0.050% | 0.051% | 0.051% | 0.077% | 0.074% | 0.077% | 0.080% |
| H_P5+_A1_€3_EV0_HP0 | 0.074% | 0.086% | 0.088% | 0.094% | 0.054% | 0.051% | 0.051% | 0.051% | 0.092% | 0.086% | 0.087% | 0.088% |
| H_P5+_A2_€1_EV0_HP0 | 0.073% | 0.086% | 0.090% | 0.100% | 0.048% | 0.050% | 0.050% | 0.050% | 0.079% | 0.084% | 0.087% | 0.089% |
| H_P5+_A2_€2_EV0_HP0 | 0.073% | 0.085% | 0.090% | 0.102% | 0.051% | 0.050% | 0.051% | 0.051% | 0.087% | 0.085% | 0.087% | 0.090% |
| H_P5+_A2_€3_EV0_HP0 | 0.072% | 0.077% | 0.081% | 0.080% | 0.053% | 0.051% | 0.051% | 0.051% | 0.089% | 0.080% | 0.080% | 0.082% |
| H_P5+_A2_€3_EV0_HP1 | 0.116% | 0.160% | 0.194% | 0.279% | 0.055% | 0.058% | 0.054% | 0.051% | 0.177% | 0.204% | 0.196% | 0.190% |
| H_P5+_A3_€1_EV0_HP0 | 0.087% | 0.107% | 0.120% | 0.160% | 0.050% | 0.051% | 0.051% | 0.050% | 0.110% | 0.116% | 0.118% | 0.119% |
| H_P5+_A3_€2_EV0_HP0 | 0.082% | 0.098% | 0.107% | 0.128% | 0.050% | 0.050% | 0.050% | 0.051% | 0.101% | 0.099% | 0.102% | 0.105% |
| H_P5+_A3_€3_EV0_HP0 | 0.077% | 0.085% | 0.092% | 0.100% | 0.052% | 0.050% | 0.050% | 0.051% | 0.096% | 0.089% | 0.090% | 0.092% |
| H_P5+_A3_€3_EV0_HP1 | 0.120% | 0.166% | 0.202% | 0.286% | 0.056% | 0.059% | 0.054% | 0.050% | 0.187% | 0.217% | 0.205% | 0.196% |
| H_P5+_A3_€3_EV1_HP0 | 0.108% | 0.157% | 0.221% | 0.460% | 0.044% | 0.043% | 0.043% | 0.045% | 0.125% | 0.126% | 0.141% | 0.157% |

## A.3. Income redistribution across IPP and DCPP

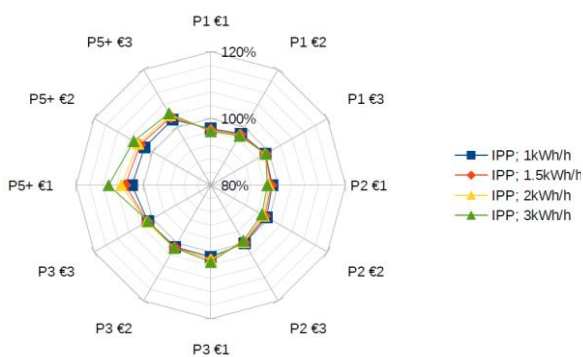

*Figure A15. Average relative change in cost with IPP for households without EV and HP across all dwelling sizes and types.*

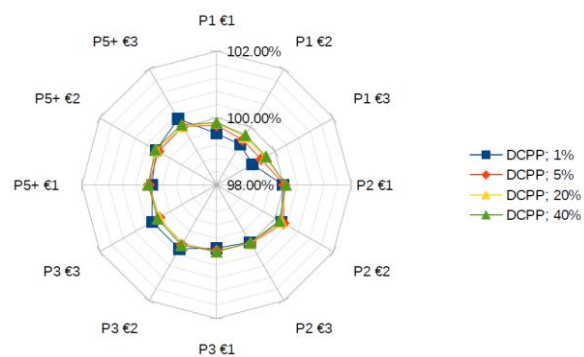

*Figure A16. Average relative change in cost with DCPP for households with EV and HP across all dwelling sizes and types.*

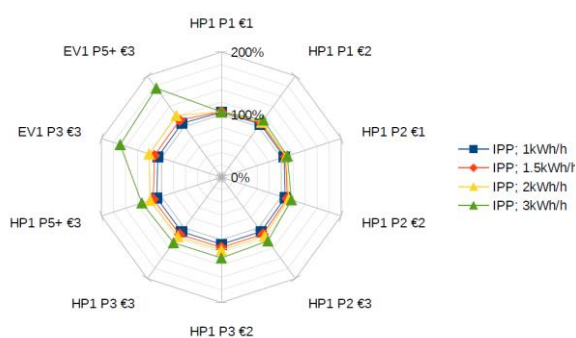

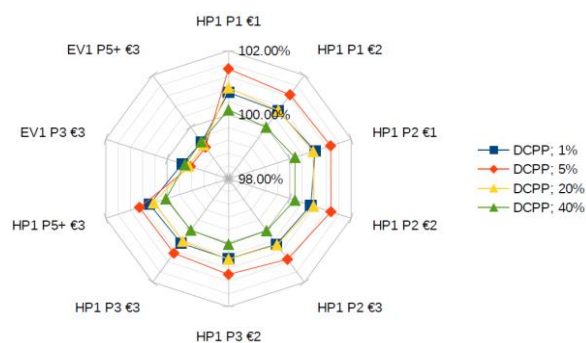

*Figure A17. Average relative change in cost with IPP for households with EV and HP across all dwelling sizes and types.*

*Figure A18. Average relative change in cost with DCPP for households with EV and HP across all dwelling sizes and types.*

**Discussion of temporal characteristics and marginal cost steps of individual peak contribution**

The IPP tariff can avoid higher costs for basic electricity consumption, as it only charges higher tariffs for peak consumption beyond a set threshold (e.g. for electricity-intensive applications: EVs, HPs). This tariff mainly targets the power/energy contribution of each consumer. However, it does not consider the temporal component to utilize the available capacities. The marginal cost around the threshold incentivizes keeping consumption below the threshold. When exceeding the threshold, consumers pay peak tariffs on their entire demand during that time. The marginal grid cost of increasing the demand by one unit stays constant regardless of whether they consume just over the threshold or significantly more. In economic terms, the marginal cost of increasing consumption when exceeding the threshold is constant. One possibility for achieving a smoother and increasing marginal cost curve is to apply multiple thresholds with increasing peak tariffs. Regardless of the detailed design of the IPP, investments in technologies that allow for a net baseload and flattened consumption are incentivized in the long run in this scheme.

In contrast to IPP, DCPP is well suited to accurately reflecting grid congestion, regardless of consumers' individual contributions. Dynamic prices are a well-suited economic tool with a constant marginal cost during peak hours, considering that peak hour forecasting develop in their temporal fragmentation towards real time operation. The constant marginal cost incentivizes consumers to reduce their entire consumption as much as possible when the network is congested. Marginal cost changes across the temporal dimension and thus induces an average effect that makes consumers invest in technologies that can shift loads across time. Despite the strong positive characteristics in terms of targeting congestion, DCPP comes with potential shortcomings in terms of tariff design objectives such as cost recovery and allocation, reflectivity, and non-discrimination (European Union Agency for the Cooperation of Energy Regulators, 2021).